\documentclass[12pt]{article}
\usepackage{epsfig, graphics}
\usepackage{amsmath,amssymb}
\usepackage{subcaption}
\usepackage[latin1]{inputenc}
\usepackage{graphicx}
\usepackage{epsfig}
\usepackage{amsfonts}
\usepackage{mathtools, bm}
\usepackage{amssymb, bm}
\usepackage[latin1]{inputenc}

\def\bkR{{\rm I\kern-.17em R}}

\newcommand{\nin}{\noindent}
\def \id{1\hskip -3pt \mbox{l}}
\def \1n{1\hskip -3pt \mbox{N}}

\def \Frac {\displaystyle \frac }
\def \Int {\displaystyle \int }
\def \Sum {\displaystyle \sum }

\newfont{\bbf}{cmbx12 scaled 1435}

\begin{document}

\thispagestyle{empty}

\vspace*{0.5cm}
\begin{center}

\setlength{\baselineskip}{.32in}

{\bbf Long Run Risk in Stationary Structural Vector Autoregressive Models}

\vspace{2cm}

\large{C.,  GOURIEROUX}\footnote[1]{Toronto University,  Toulouse School of Economics, and CREST {\it e-mail}:
{\tt christian.gourieroux@ensae.fr}} and
\large{J.,  JASIAK }\footnote[2]{York University, Canada, {\it e-mail}:
{\tt jasiakj@yorku.ca}.  \\
The first author gratefully acknowledges financial support from the ACPR chair "Regulation and Systemic Risk", and the ERC DYSMOIA. The second author thanks the Natural Sciences and Engineering Council of Canada (NSERC).}

\vspace{1cm}
February, 16, 2022

\bigskip
\begin{minipage}[t]{12cm}
\small
\begin{center}

Abstract\\
\end{center}

This paper introduces a local-to-unity/small sigma process for a stationary time series with strong persistence  and non-negligible long run risk. This process represents the stationary long run component in an
unobserved short- and long-run components model involving different time scales. More specifically, the short run component evolves in the calendar time and the long run component evolves in an ultra long time scale.  We develop the methods of estimation and long run prediction for the univariate and multivariate  Structural VAR (SVAR) models with unobserved components and reveal the impossibility to consistently estimate some of the long run parameters. The approach is illustrated by a Monte-Carlo study and an application to macroeconomic data.


\end{minipage}
\end{center}

\renewcommand{\thefootnote}{\arabic{footnote}}
\setcounter{page}{1}

\newpage
\section{Introduction}
\setcounter{equation}{0}\def\theequation{1.\arabic{equation}}

The macroeconomic and financial models  used in practice provide reliable predictions at short horizons of 1 to 5 years. Recently, there has been a growing interest in providing long run predictions at horizons of 10 to 50 years, in the context of transition to low carbon economy,  climate risk and rare extreme events, for evaluating of necessary behavioral and technical changes. In the financial sector as well, the long run predictions may soon become mandatory for prudential supervision in banks and insurance companies.

The long-run predictions are difficult to compute for several reasons. The standard prediction models with short lags and the associated statistical inference methods are inadequate for long horizons. Moreover, long run predictions at horizons of 50 or 100 years are difficult to compute from macroeconomic or financial time series observed over periods  shorter or equal to the prediction horizon of interest. Therefore, the long run predictions remain mainly model based. As such, they can be improved so that:

i) The estimation methods account for the long run component, even though it is difficult to detect over the sampling period.

ii) The long run predictions produce reasonable outcomes in the sense that point forecasts should take values from the set of admissible values of the predicted variable.

iii) The prediction errors are not underestimated due to the selected dynamic model and its estimation method.

This paper examines the feasibility of such improvements in the class of dynamic stochastic linear models with long run properties already known in the literature to some extent. The Structural Vector Autoregressive (SVAR) models are stochastic linear systems which are second-order identifiable under a set of parameter restrictions. The SVAR models are used in macroeconomics, monetary economics and macro-finance for the analysis of multivariate economic processes and prediction of future shock effects through impulse response functions [see Hurwicz (1962) for the introduction of the term "structural" and Sims (1980), (2002) for SVAR applications]. Two types of SVAR models can be distinguished, which are the stationary (regular) SVAR models and cointegrated SVAR models with nonstationary unit root components. While both types of SVAR models are efficient instruments of short run analysis, they have limitations in application to the long run and ultra long run analysis. More specifically,

i) the presence of nonstationary features, such as nonstationary unit roots, implies explosive patterns in trajectories, which are incompatible with the behavior of variables such  as the growth rate of per capita real GDP, productivity, real food expenditure per capita, interest rates, real exchange rates, and some commodity prices, or spot-forward spreads. \footnote{See for instance the discussion on the mean reverting feature of consumption in Beeler, Campbell (2012) and the analysis of spot-forward spreads in Gospodinov (2009), Gospodinov et al. (2021).} In general, variables transformed to stationarity by differencing or expressed as rates of growth are not explosive; ii) the identification restrictions imposed on the long run behaviour, i.e. the so-called long run identification restrictions  [Blanchard, Quah (1989), Christiano et al. (2006)] can affect the long-run predictions and the long run patterns of impulse response functions [Sims (1986)] as well as their signs [Uhlig (2005)]; iii) the long run risk is diversified away and often disregarded in stationary SVAR models
[see e.g. Bansal, Yaron (2004), Bansal, Kiku, Yaron (2010), Croce, Lettau, Ludvigson (2015) for recent papers on the long run analysis with regular SVAR models estimated as stationary processes].

Among the macroeconomic time series, those considered stationary often display persistence up to high lags, in the sample autocorrelation functions. Therefore, a large body of literature use the local-to-unity model linking the aforementioned two types of SVAR dynamics and develop associated inference methods [see e.g. Chan, Wei (1987), Phillips (1987), Magdalinos, Phillips (2007), and the recent survey by Muller, Watson (2020)]. However, the standard local-to-unity models may not be suitable for stationary time series (see the discussion in Appendix 1).

The aim of our paper is to propose an alternative approach by considering a stationary VAR model with a stationary short run (SR) component and a stationary multivariate ultra long run (ULR) component with asymptotic unit root and close to zero sigma,

For illustration, let us consider the framework of stationary Gaussian processes (where strict stationarity and second-order stationarity are equivalent). This simplified framework is not only convenient for comparing our analysis with the large body of literature on structural SVAR models, but also for introducing and interpreting the notion of an ultra long run process in a stationary time series seen as a sum of a regular and a singular components. There exist two generic representation theorems of a time series seen the Wold representation theorem for weakly stationary processes [Wold (1938)] and the Volterra representation theorem for strictly stationary processes [Rozanov (1967), Priestley (1988)]. These two representations coincide for the Gaussian stationary processes. According to the Wold representation,  any (multivariate) stationary Gaussian process $y(t)$ can be written as:

\begin{equation}
  y (t) = \sum_{j=0}^{\infty } A_j \epsilon_{t-j} + Z_t,
\end{equation}

\noindent where $dim \, y = \dim \epsilon = dim \; Z \equiv n$, the $(n \times n)$ matrices $A_j$ are such that $\sum_{j=0}^{\infty} A_j A_j'$ exists and is finite, $(\epsilon_t)$ is a Gaussian white noise, $\epsilon_t \sim IIN(0, Id)$. The moving average component is the regular component.
The component $Z_t$ is the singular component of process $(y(t))$ and is measurable with respect to
$F_{-\infty} = \cap_t F_t$ where $F_t$ is the $\sigma$-algebra generated by the current and lagged values of $y$.
Following Doob (1944), this singular component is also referred to the deterministic component of the process. This explains why in
practice, $Z_t$ is often disregarded and replaced by a deterministic constant:

\begin{equation}
  y (t) = \sum_{j=0}^{\infty } A_j \epsilon_{t-j} + m.
\end{equation}

\noindent The term "deterministic" is misleading, because the stationary singular component can be constant over time while being stochastic, so that $Z_t=Z$ is independent of the noise and normally distributed: $Z \sim N(m, \Sigma)$, where $\Sigma$ does not necessarily degenerate to zero.
The fact that $Z_t$ is measurable with respect to $F_{-\infty}$ means that it is only influenced by the infinitely distant past and has a long run interpretation. In particular, the singular component is such that:

$$E[Z_{t+h}| F_{t}] = Z_t = Z, \; \mbox{for any } h,t. $$

\noindent Thus $Z_t$ is a stationary martingale, which is a particular  \textbf{stationary unit root process} \footnote{Since $Z_{t+h} = Z_t =Z$, the theoretical prediction error is equal to zero. This explains the terminology "purely predictable" process also used for this singular component. This other terminology is also misleading since in practice we observe $y$ from an initial date only, and, if $F_{0,t}$ is the $\sigma$-algebra generated by $y(0), \ldots, y(t)$, then $E (Z_{t+h}| F_{0,t})$ is different from $Z_t$.}.
The singular component can  be  approximated by a moving average:

$$
Z_T(t) = \sum_{j=0}^{\infty } A_{Tj}^* \epsilon^*_{t-j},
$$

\noindent where $(\epsilon^*_t)$ is another Gaussian white noise, independent of $(\epsilon_t)$ and $\Sum_{j=0}^{\infty } A_{Tj}^* A_{Tj}^{*'}$ is finite. This approximation leads to the triangular array (doubly indexed by $t$ and $T$):

\begin{eqnarray}
y_T(t) & = & \sum_{j=0}^{\infty } A_j \epsilon_{t-j} + \sum_{j=0}^{\infty } A_{Tj}^* \epsilon^*_{t-j} \nonumber \\
& = & \sum_{j=0}^{\infty } A_j \epsilon_{t-j} + Z_T(t),
\end{eqnarray}

\noindent which approximates the representation (1.1) of $y(t)$.

The stationary moving average approximation of the singular component $Z_t=Z$ is slowly varying over time and has a negligible effect on the dynamics of $y(t)$ in the short run. As shown later in the text, it has a significant impact in the long run dynamic, and is therefore interpreted as the Ultra Long Run (ULR) component. Moreover, process $\left\{ y_T(t) \right\}$ remains stationary, but becomes non-ergodic.

More generally, in the VAR framework we can write the n-dimensional process $y_T(t)$ as a sum of the short run (SR) component $y_s(t)$, which for ease of exposition is assumed to follow a VAR(1) process, and an ULR component:

\begin{equation}
   y_T(t) = y_s(t) + A y_l (t/T),
\end{equation}

\noindent where $A$ is a $(n,L)$ matrix of coefficients of full column rank L and :

\begin{equation}
  y_s(t) = \Phi y_s(t-1) + \Omega^{1/2} \epsilon_t,
\end{equation}

\noindent where the eigenvalues of $\Phi$  are of modulus strictly less than 1 to ensure the stationarity of the SR process $y_s$, the $(n,n)$ matrix  $\Omega$ is positive definite and $\epsilon_t$ is a multivariate strong white noise process with second-order moments, but not necessarily Gaussian. The component $[y_l (\tau)]$ follows a L-dimensional (continuous time) multivariate Ornstein-Uhlenbeck process:

\begin{equation}
  d y_l(\tau) = -\Theta y_l(\tau) d \tau + S dW_{\tau},
\end{equation}

\noindent where $\Theta$ is of dimension $L \times L$ and such that the eigenvalues of matrix \footnote{$\exp (-\Theta)$ is defined as~: $\exp (-\Theta) = \sum^\infty_{j=0} [(-1)^j \Theta^j/j!].$} $\exp (-\Theta)$ are of modulus strictly less than one to ensure the stationarity of process $y_l$. Process  $(W_{\tau})$ is a L-dimensional Brownian motion independent of the noise in the short run component and $S$ is a positive definite matrix of dimension $(L,L)$. The ULR component is deduced from the continuous time underlying process $[y_l (\tau)]$ by applying the time deformation and time discretization. More specifically, the short run component evolves over calendar time, whereas the long run component evolves over time measured on a scale with an ultra large time unit $T$ that tends to infinity with $T$. In particular, for any date $t$, $y_l(t/T)$ will tend to $y_l(0)$, which is constant over time and random. This limit is a singular process.
Equations (1.4)-(1.6) impose a special structure on the VAR(1) model, turning it into a Structural VAR where the SR and ULR shocks can be identified.

The process (1.4)-(1.6)  is an example of a local-to-unity model with behavioral foundations [see the discussion in Sims (1988), Section 2]. Indeed, its dynamics is compatible with a dynamic equilibrium of two types of individuals (consumers, investors), who operate at two different frequencies and have different information sets.
By considering these two frequencies jointly, we do not support the "belief that data collection and modelling strategy usually lead economists to use data corresponding to a "fine" time unit relative to the phenomena they are studying" [Sims (1991), p429].

In the approximate Wold representation (1.3) of (1.4)-(1.6)  the number of underlying shocks is equal to $n+L = dim \epsilon_t + dim W_{\tau}$, which is larger than the dimension of $[y_T(t)]$. These two types of shocks are measured on different time scales. The number of ULR processes $L$ is assumed to be less that $n$ (the dimension of $Z$) by imposing the condition of full column rank on matrix $A$. The $(n,L)$ matrix  $A$ is introduced to identify the components of the array that are not sensitive to long run shocks. More precisely, for any direction $\gamma$ such that $\gamma'A=0$, the combination $\gamma ' y_T(t) = \gamma' y_s(t)$, no longer depends on the ULR component and $T$. This allows us for considering the ULR co-movement relationships (i.e. the $\gamma' s)$.

The paper is organized as follows. Section 2 discusses the second-order properties of the model such as the theoretical autocovariances and spectral density of the (triangular) array. We show how the theoretical autocovariances and spectra depend on $T$, derive their limits when $T$ tends to infinity and reveal a lack of uniformity in their convergences. As a consequence, new identification issues arise for statistical inference.

Statistical inference is discussed in Section 3. We focus on estimation methods based on the  first- and second-order sample moments. Under the triangular array representation $[y_T (t)],$ the sample autocovariances can have asymptotic properties which are very different from the standard properties. Therefore, the  sample autocovariances and autocorrelations may not be fully informative about the parameters of interest. We show that different types of autocorrelation functions (ACF) have to be considered jointly. These are:

i) the standard ACF evaluated at multiples of large lags;

ii) short run ACF evaluated at short lags over short episodes of time;

iii) long run ACF at large lags computed from historical averages of the observed process $[y_T(t)]$.

\noindent We show that, from rolling averages and different ACF's, it is possible to estimate consistently (under the standard structural identification restrictions for SVAR):

i) the parameters $\Phi$ and $\Omega$ driving the short run dynamics;

ii) the levels $[y_l(\tau), \tau \in (0,1)]$ of the underlying long run continuous time

\hspace{0.5cm} model;

iii) the matrix A of factor sensitivities;

iv) the volatility of the ULR component, i.e. the matrix $SS'$.

The "estimators" in i)-iv) converge at different rates. In particular the ULR parameters converge at slower rates, which reveals weak identification issues. However, the main identification  issue is the impossibility to estimate consistently the drift $\Theta$ in the Ornstein-Uhlenbeck (OU) dynamics (the \textbf{impossibility theorem}). This implies that, even with an infinite number of observations, the long run estimation risk cannot be diversified away. Thus, the evaluation of the long run estimation risk is crucial for the long run predictions and impulse responses.
Section 4 highlights the importance of taking into account the estimation risk when computing the long run prediction intervals and the Value-at-Risk. We also discuss the prudential principle and the limitation of Bayesian approaches in this respect. Section 5 illustrates the proposed methodology by Monte-Carlo studies and provides an empirical application to a set of macroeconomic series. It compares the approach introduced in this paper with the treatment of long run risk in the applied macroeconomic literature. Section 6 concludes. A comparison with alternative local-to-unity models is provided in Appendix 1. The Law of Large Numbers (LLN) for triangular arrays used for deriving the asymptotic results is given in Appendix 2, and the Central Limit Theorems (CLT) in Appendix 3. Additional figures are provided in Appendix 4.

\section{Dynamic properties}

\subsection{The VAR representation}
\setcounter{equation}{0}\def\theequation{2.\arabic{equation}}

A time discretized multivariate Ornstein-Uhlenbeck process is a Gaussian VAR(1) process. Hence, the triangular array model (1.4) - (1.6) can be rewritten as [see Meucci (2010)]:

\begin{equation}
y_T(t) = y_s(t) + A y_{l,T}(t),
\end{equation}

\noindent where~:

\begin{equation}
y_s(t) = \Phi y_s (t-1) + \Omega ^{1/2} \epsilon_t,
\end{equation}

\begin{equation}
y_{l,T}(t) = exp(-\Theta/T)y_{l,T}(t-1)  + \Sigma_T ^{1/2} \epsilon_{t}^*,
\end{equation}

\noindent  with~:

\begin{equation}
vec(\Sigma_T) = [ \Theta \oplus \Theta]^{-1} \{ Id - exp [-(\Theta \oplus \Theta)/T]\} vec (SS'),
\end{equation}

\noindent and the Kronecker sum defined by:
\begin{equation}
\Theta \oplus \Theta = Id \otimes \Theta + \Theta \otimes Id,
\end{equation}

\noindent with $\otimes$ the Kronecker product.

System (2.1) -(2.4) shows that the process $[(y_T(t)]$ for any given $T$ is the sum of two independent VAR(1) processes \footnote{When the unobserved components are integrated out, the process $[y_T (t)]$ follows a weak VARMA model with orders depending on $L$ and on the assumptions on $\Phi$ and $\Theta$. This interpretation is not used later on.}.  These two processes are stationary, when their initial values are drawn in a stationary distribution such as, for example,  N(0, $\Sigma$), with $vec \Sigma = [ \Theta \oplus \Theta]^{-1} vec (SS')$ for the ULR component.

The VAR dynamics of the ULR component is such that the autoregressive coefficient $exp(-\Theta/T)$ tends to $Id$, when $T$ tends to infinity while at the same time the volatility $\Sigma_T$ given in (2.4) tends to 0. Therefore it is a sequence of multivariate stationary processes, with an asymptotic multivariate unit root and close to zero sigma. They are examined
in the framework of a local-to-unity and small sigma analysis with respective rates of convergence to Id and 0 being of equal order \footnote{Since $\exp (-\Theta/T) \sim Id-\Theta/T, \exp [-\Theta \oplus  \Theta/T] \sim Id - (\Theta \oplus \Theta)/T.$} $1/T$ [see Kadane (1971) for the introduction of small sigma asymptotics to econometrics]. This approach differs from the local-to-unity analysis with fixed sigma that leads to nonstationary processes at the limit [see e.g. Stock (1991), p 437, Muller, Watson (2016), p 1723, in the univariate case]. To clarify this difference with the literature (we elaborate on this point in  Appendix 1), let us consider below  a univariate process:

\begin{equation}
y_T(t) = y_s(t) +  y_{l,T}(t),
\end{equation}

\noindent where~:

\begin{equation}
y_s(t) = \phi  y_s(t-1) + \eta \varepsilon_t, \eta> 0,
\end{equation}

\begin{equation}
y_{l,T}(t) = exp(-\theta/T) y_{l,T}(t-1) + [ \frac{s^2}{2\theta} [1-\exp(- 2 \theta /T)]]^{1/2} \epsilon_{T,t}^*.
\end{equation}

For a given $T$, the variance of the innovation of the ULR component is proportional to $1-\rho_T^2$, where $\rho_T = exp(-\theta/T)$ is the autoregressive coefficient. Therefore, the autoregressive coefficient and the volatility tend to 1 and 0, respectively, in a constrained manner that is consistent with the time deformation interpretation.

An array with such a dynamic might be interpreted as a local level model. Let us recall the standard  definition of a local level model introduced by Harvey (1989) [see also Durbin, Koopman (2002)]. In our notation, a local level model is a time series model such that:

$$y(t) = y_s(t) + y_l(t), $$

\noindent where $y_s(t) = \eta \epsilon_t$, $ y_l(t) =  y_l(t-1) + s \epsilon_t^*$. Like our model, the local level involves a decomposition with a short term component, here a noise process, and a long run component, which is a random walk. The main difference between the standard  local level and our model is the specification of the long run component. The random walk representing the long-run component in the standard local level  model is nonstationary and has "explosive" trajectories in the long run unlike the ULR process (see also the discussion in Appendix 1 of the different local-to-unity models proposed in the literature).

As shown in the next Section, the ULR model allows for a stochastically varying "local level" without an explosive pattern. The ULR model also differs from the model with a local level smoothly varying in time in a deterministic pattern [Dalhaus (2012), Dalhaus et al. (2017)].

\subsection{Autocovariances and Spectrum}

Let us now consider the theoretical autocovariances and spectrum. For ease of exposition, we consider the univariate model (2.6)-(2.8) and known parameter values.

\noindent  For a fixed $T$, the autocovariance of process $y_T(t)$ defined by (2.6)-(2.8) at lag $h$ is given by:

\begin{equation}
\gamma_T(h) = \eta^2 \varphi^h + \frac{s^2}{2 \theta} [ 1- exp( - \frac{2h \theta}{T})], \; h \geq 0.
\end{equation}

\noindent We have the following proposition:

\medskip

\noindent {\bf Proposition 1}\vspace{1em}

i) If $h$ is fixed, $T \rightarrow \infty$, $\gamma_T(h) \rightarrow \eta^2 \varphi^h.$

ii) If $h=h_T \rightarrow \infty$, when $T \rightarrow \infty$, $\gamma_T(h_T) \sim \frac{s^2}{2 \theta} [ 1- exp( - \frac{2h_T}{T})]$.\vspace{1em}

\noindent In particular,
if $h_T \rightarrow \infty$, $T \rightarrow \infty , h_T/T \rightarrow 0$, then $\gamma_T(h_T) \rightarrow 0$.\vspace{1em}

\noindent If $h_T \rightarrow \infty$, $T \rightarrow \infty , h_T/T \rightarrow c$, then $\gamma_T(h_T) \rightarrow
\frac{s^2}{2 \theta} [1-exp(-2 c \theta)]$.\vspace{1em}

\noindent The asymptotic behaviour of the sequence of autocovariances is complex as it is different at short and ULR lags. For the short lag (resp. long lag) only the short run (resp. long run) component matters. Moreover, the convergence of this autocovariance function to a limit is not uniform. This causes identification issues discussed in the next Section.

Similarly, let us consider the spectrum at a given date $T$ given by~:

\begin{equation}
S_T(w) = \frac{\eta^2}{2\pi} \frac{1}{1+\varphi^2 - 2 \varphi cos w} + \frac{s^2}{2 \pi \theta} \frac{1-exp(-2 \theta/T)}{1 + exp (-2 \theta/T) - 2 exp (-\theta/T) cos w}.
\end{equation}

\noindent We get the following limiting behaviour:

\medskip

\noindent {\bf Proposition 2}

i) For $w$ fixed, $T \rightarrow \infty$, $S_T (w) \rightarrow
\frac{\eta^2}{2 \pi} \Frac{1}{1 + \varphi^2 -2 \varphi \cos w} $, if $ w\neq 0$,

$S_T (w) \rightarrow \infty$ , if $ w=0$.

ii) If $w_T \rightarrow 0^+$, $T \rightarrow \infty$,

\begin{eqnarray*}
S_T(w_T) & \sim & \frac{\eta^2}{2 \pi} \frac{1}{(1-\varphi)^2} + \frac{s^2}{2 \pi \theta}  \frac{1-exp(-2 \theta/T)}{1 + exp (-2 \theta/T) - 2 exp (-\theta/T) cos w_T} \\
& &\sim \frac{\eta^2}{2 \pi} \frac{1}{(1-\varphi)^2} + \frac{2 \theta/T}{w_T^2}.
\end{eqnarray*}

\noindent In particular, if  $T w^2_T \approx \lambda$, then  $S_T (w_T) \approx \Frac{\eta^2}{2\pi} \Frac{1}{(1-\varphi)^2} + \Frac{2 \theta}{\lambda}$.\vspace{1em}

The lack of uniform convergence of the sequence of spectra is analogous to the lack of uniform convergence of the sequences of autocovariances discussed above. For a fixed $w$, $w \neq 0$, the limit depends on the short run dynamics only. When $w_T$ tends to 0 at an appropriate rate $1/\sqrt{T}$, the limit involves both the short- and  long-run parameters, $\varphi$ and $\theta$, respectively.

\section{Statistical Inference}
\setcounter{equation}{0}\def\theequation{3.\arabic{equation}}

So far, we have discussed the asymptotic behaviours of the theoretical autocovariances and spectrum in Section 2.2. This Section examines the asymptotic properties of the sample autocovariances and sample spectrum computed from the observations on $y_T(t), t=1,...,T,$ corresponding to a triangular array.

The standard sample autocovariances are based on global averages that conceal the local mean effect due to the ULR component. Therefore, we distinguish the three following types of sample ACF:

i) the standard ACF evaluated at long lags from observations that are sufficiently far apart;

ii) the short run ACF computed at short lags from short time series of observations;

iii) the long run ACF based on short run averages of observations.

\noindent As shown below, these ACF's have distinct behaviours and interpretations, explaining the long run predictability puzzle reported in Fama, French (1988), (1989), or the debate on the predictability of consumption growth from log price/dividend ratio [Becler, Campbell (2012), Section 4.1]. This shows also that the standard asymptotic properties of moment estimators have to be modified in models with long run component [see e.g. Bansal, Kiku, Yaron (2012), (2016) for the use of GMM in models with long run risk].

\subsection{ACF Evaluated at Distant Lags}

\noindent Let us denote the multivariate sample autocovariance by:

\begin{equation}
\hat{\Gamma}_T (h)  = \frac{1}{T-h} \sum_{t=h+1}^T y_t y_{t-h}' -  \frac{1}{T-h} \sum_{t=h+1}^T \; y_t \; \frac{1}{T-h} \sum_{t=h+1}^T y_{t-h}'.
\end{equation}

\noindent From Proposition 1 ii) given for the one-dimensional case, we expect a rather regular behaviour of the multivariate sample autocovariances at long lags $h_T = cT$ for any given $c, c \in (0,1).$

To interpret this result, let us first consider the sample mean. We have:

$$
 \frac{1}{T-h_T} \sum_{t=h_T+1}^T y_t  =  \frac{1}{T-cT} \sum_{t=cT+1}^T y_s (t) +  \frac{A}{T-cT} \sum_{t=cT+1}^T y_l (t/T).
$$

\noindent The first term tends to 0 by the standard Law of Large Numbers applied to the short run stationary component. The second term is a Riemann sum that tends to the associated stochastic integral, by applying the Stroock, Varadhan theory of diffusion approximation [Stroock, Varadhan (1979), Section 11]. In brief, we have asymptotically:

\begin{equation}
 \frac{1}{T-cT} \sum_{t=cT+1}^T y_t \rightarrow \frac{A}{1-c} \int_c^1 y_l (v) dv,
 \end{equation}

\noindent where $\rightarrow$ denotes the convergence in probability (and also the weak convergence of processes indexed by $c$).

\noindent The sample mean does not converge to the theoretical mean of process $[y_T(t)]$ equal to 0. A similar derivation can be done for the cross-product (see Appendix 2).
This result is summarized in the Proposition below.

\medskip
\noindent {\bf Proposition 3}

\noindent For large $T$,

\begin{eqnarray*}
\hat{\Gamma}_T (cT) &  = & \frac{A}{1-c} \int_c^1 y_l (v)  y_l' (v-c) dv \; A'\\
&  & - \left[\frac{A}{1-c} \int_c^1 y_l (v) dv\right] \left[  \frac{A}{1-c} \int_c^1 y_l (v-c) dv \right]' + o_p(1),
\end{eqnarray*}

\noindent where $o_P (1)$ is negligible in probability.\vspace{1em}

Proposition 3 reveals a non-standard asymptotic behavior of sample autocovariances evaluated at long lags. When $h$ increases, these sample autocovariances do not converge to constant values, such as their theoretical analogues (i.e. $\frac{s^2}{2 \theta} [1-exp(-2 \theta)]$ in the univariate case, see Proposition 1). The sample autocovariances at large lags still converge, but to a stochastic limit that depends on the trajectory of the ULR component.

\subsection{Short Run Sample Autocovariances}

For a fixed $h$, the sample autocovariance $\hat{\Gamma}_T(h)$ does not provide a good approximation of $\Gamma_T(h),$ which is the autocovariance function of the short run component (which follows from Proposition 1, i)). Indeed, $\hat{\Gamma}_T(h)$ is a global measure that is impacted by the local level effect due to the ULR component. This effect can be adjusted for as follows. Let us consider a subset of observations corresponding to the interval $(cT , cT+H_T),$ where $H_T \rightarrow \infty, H_T/T\rightarrow 0$, and compute the sample autocovariances over this interval:

\begin{equation}
\hat{\Gamma}_{c,T} (h)  = \frac{1}{H_T} \sum_{t=cT+1}^{cT+H_T} y_t\; y_{t-h}' -
\frac{1}{H_T} \sum_{t=cT+1}^{cT+H_T} y_t \;\frac{1}{H_T} \sum_{t=cT+1}^{cT+H_T} y_{t-h}'.
\end{equation}

\noindent Because $H_T <<T$ and the ULR component is close to the purely predictable component, we have:

\begin{eqnarray*}
y_t & = & y_s(t) +  A y_l (t/T) \\
& \approx & y_s(t) +  A y_l (c) + o_p (1),
\end{eqnarray*}

\noindent for $t \in (cT+1, cT + H_T)$. The process corresponds to the short run component plus a stochastic "local mean" equal to $A y_l (c)$. Then the standard LLN can be applied to the process conditional on $[y_l (\tau)]$.

\medskip

\noindent {\bf Proposition 4}

For any fixed $h, h \leq H_T$, any $c,c \in (0,1),$ and any $H_T,$ with $ H_T \rightarrow \infty, H_T/T \rightarrow 0$, we have~: $\hat{\Gamma}_{c,T}(h) \rightarrow \Gamma_s(h),$ where $\Gamma_s(h)$ is the theoretical autocovariance function of the short run component.\vspace{1em}

The above result can be used to estimate consistently some parameters of interest. Let us consider a grid $c_1 \leq ... \leq c_{K_T}$ of $[0,1]$ .

\noindent Then, for each $c_k$ we compute the sample mean:

\begin{eqnarray}
\hat{m}_T(c_k) & = & \frac{1}{H_T} \sum_{t=c_kT+1}^{c_kT+H_T} y_t  \\
& \approx & \frac{1}{H_T} \sum_{t=c_kT+1}^{c_kT+H_T} [y_s(t) + A y_l (c_k)] \nonumber \\
& \approx & A y_l (c_k), \mbox{by the LLN applied to the short run component.}\nonumber
\end{eqnarray}

\medskip

\noindent {\bf Corollary 1:}

The sample means $\hat{m}_T(c_k)$ are consistent estimates (predictors) of the values of the underlying URL process, up to the multiplicative matrix $A$. By standard arguments, the sample means converge at speed $1/\sqrt{H_T}$.\vspace{1em}

\noindent Then several estimates of the autocovariances of the short run component $ \hat{\Gamma}_{c_k ,T} (h), k=1,...,K$ can be computed and used to estimate $\Gamma_s(h)$ by:

\begin{equation}
 \hat{\Gamma}_{T,s} (h) = \frac{1}{K}
\sum_{k=1}^{K}  \hat{\Gamma}_{c_k,T} (h), h\leq H_T.
\end{equation}

\medskip

\noindent {\bf Corollary 2:}

The autocovariances $\Gamma_s(h), h\leq H_T,$ can be consistently estimated by $\hat{\Gamma}_{T,s} (h), h\leq H_T$. They converge at speed $1/\sqrt{H_T K}$.\vspace{1em}

In a structural SVAR model, under the usual identification restrictions on $\Phi, \Omega$, consistent estimates of the identifiable short run parameters can be obtained from the moment conditions:

 \begin{eqnarray}
\Gamma_s(1)  =  \Phi \Gamma_s(0),\;\;\; \Gamma_s (0)   =  \sum^\infty_{j=0} \Phi^j \Omega \Phi^{j'},
 \end{eqnarray}

\noindent after replacing $\Gamma_s(h), \; h=0,1,...$ by $\hat{\Gamma}_{T,s}(h)$ These estimators converge at speed $1/\sqrt{H_T K}$.\vspace{1em}

Let us now discuss the estimation of matrix $A$. A multivariate factor model is commonly defined up to an invertible transformation. In the linear framework, for any $(L,L)$ invertible matrix $Q$, we get the same  $L$-variate model with $A$ replaced by $A Q^{-1}$ and $y_l$ replaced by $Q y_l$ that is an O.U. process as well. Under the identification restriction on $A$ which is imposed to solve this multiplicity, we get an estimate $\hat{A}_T$ of $A$ by using the principal component analysis (PCA). Next, by applying a pseudo inverse of $\hat{A}_T$, which is $\hat{A}^+_T = (\hat{A}'_T \hat{A}_T)^{-1} \hat{A}'_T,$ approximations $\hat{y}_{lT} (c_k) = \hat{A}^+_T \hat{m}_T (c_k)$ of $y_l (c_k)$ are obtained. This leads to the following corollary: \footnote{If the identification restriction on $C$ is $C' C = Id_L$, the pseudo inverse becomes $\hat{C}^+_T = \hat{C}'_T$.}

\medskip

\noindent {\bf Corollary 3:}

The matrix $A$ and the values $y_l (c_k)$ are consistently estimated with the convergence rate of estimators of $1/\sqrt{H_T}$.\vspace{1em}

\noindent In particular,

\medskip

\noindent {\bf Corollary 4:}

For $T = \infty$, the long run component $y_l (c), c \in [0,1],$ is filtered without errors. Since

$$
y_s(t) = y_t - y_l(t/T) \approx y_t - \hat{y}_{lT} (c),
$$

\noindent for $t \in (cT, cT + H_T)$, the short run component is  known as well in any neighbourhood of a date $cT$.\vspace{1em}

Corollary 4 shows that asymptotically, the underlying  SR and ULR components can be detected. Therefore, in this model involving two time scales, the identification of the short and long run dynamics at different "frequencies" is possible without introducing any additional structural identifying restrictions. This result is analogous to the asymptotic detection of smooth deterministic local level [Dalhaus (2012), Dalhaus et al. (2017)], although in a context of stochastic local levels. Corollary 4 is also compatible with the remark of Beeler, Campbell (2012) in a critic of the Bansal et al. (2012) long run modelling approach~: "Predictable variations in long run (component) might exists in the data, but might be masked by temporary fluctuations that are omitted from the long-run risk model. If this is the case, however, economic agents must perceive those variations in order for them to influence asset prices".

It is also possible to derive the asymptotic distribution of the difference between the local sample mean $\hat{m}_T (c_k)$ and its stochastic limit $m(c_k) = A y_l (c_k)$ (see Appendix 3).\vspace{1em}

\textbf{Proposition 5~:} Let us assume that $H_T \rightarrow \infty$ and $H_T^2/T \rightarrow 0$, when $T\rightarrow \infty$. Then we have~:

$$
\sqrt{H_T} [ \hat{m}_T (c_k) - A y_l (c_k), k=1,\ldots, K]
$$

 \noindent tends in distribution to $N (0, Id \otimes \Sigma_\infty),$ with~:

$$
\Sigma_\infty = \sum^{+\infty}_{h=-\infty} \;\mbox{cov}\; [y_s (t), y_s (t-h)].
$$

This asymptotic result can be used to derive the asymptotic distributions of $\sqrt{H_T} [\hat{A}_T - A]$ and of $\sqrt{H_T} [\hat{y}_{lT} (c_k) - y_l (c_k)]$.

Proposition 5 shows that the condition $H_T^2/T \rightarrow 0,$ if $T\rightarrow \infty$, eliminates the effect of the smooth varying stochastic level, when deriving these asymptotic distributions. If $H_T^2/T \rightarrow  \gamma >0$, say, an additional term should have to be included in the asymptotic variance formula.

\subsection{Long Run ACF}

The dynamics of the ULR component can be analysed by means of the proxies $\hat{y}_{lT} (c)$ of $y_{l} (c)$. Asymptotically, this is as if the underlying continuous time O.U. process were observed on the interval $[0,1]$. Then, a maximum likelihood method, that is an OLS estimator in our Gaussian case, can be applied after substituting the proxies $\hat{y}_{lT} (c)$ for the true value $y_l (c)$ over the grid of $(c_1, \ldots, c_{K})$. Let us examine the consequences of disregarding the first step estimation error, that is the fact that $\hat{y}_{Tl} (c) \neq y_l (c)$ and computing the ML estimator on a regularly spaced grid which is getting finer with $T$. The following well-known result is  commonly used in the high frequency data literature.

\medskip
\noindent {\bf Proposition 6}

Let us consider a regular grid $c_1 <\ldots < c_{K_T}$ of [0,1], and assume observable $y_l (c_1), \ldots, y_l (c_{K_T})$, then if $K_T \rightarrow \infty$,

i) The volatility parameter $S$ can be estimated consistently with a convergence rate $1/\sqrt{K_T}$.

ii) The "drift" parameter $\Theta$ cannot be consistently estimated.\vspace{1em}

\noindent The last statement follows from the \textbf{impossibility theorem} for continuous time models [see Banon (1978), Jiang, Knight (1997)]. Although the drift parameter can be estimated, the distribution of its estimator does not degenerate to a point mass at the true value even when $T \rightarrow \infty$. The estimation risk on some of the long run parameters persists indefinitely, and it cannot be diversified away given the observations on a single time series. Below, we show that finite sample techniques can be used to address this issue.

\subsection{Summary of the Estimation Approach}

Under the identification restriction on the short run structural VAR and the identification restriction on the long run factor (and $A$), the parameters can be estimated along the following steps:\vspace{1em}

\textbf{Step 1~:} Compute the short run sample autocovariances at lags 0 and 1 from formula (3.5) and apply a method of moment estimation based on the moment conditions (3.6) under the short run identification restrictions on the short run SVAR. This step provides consistent estimates of short run parameters $\Phi, \Omega$.\vspace{1em}

\textbf{Step 2~:} Compute the short run local means at different dates $c_k T, k=1,\ldots,K$, from formula (3.4). This step provides approximations $\hat{m}_T (c_k)$ of $A y_l (c_k), k=1,\ldots,K.$\vspace{1em}

\textbf{Step 3~:} Concatenate the vector of $K$ summary statistics $\hat{m}_T (c_k), k=1,\ldots,K$, each of dimension $n$ to perform a principal component analysis (PCA) of the matrix~: $\Sum^{K}_{k=1} [(\hat{m}_T (c_k) \hat{m}'_T (c_k)] \equiv \hat{M}_T.$. Keep only the principal directions associated with the statistically significant eigenvalues of $\hat{M}_T$. Under the identification restriction $A'A=Id_L$ for$A$, this step provides~:\vspace{1em}

i) an estimate $\hat{L}_T$ of the number of underlying long run components;\vspace{1em}

ii) an estimate of the "factor sensitivity" matrix $A, \hat{A}_T$, say;\vspace{1em}

iii) the approximations of the underlying long run component values at the $c'_k s, k=1,\ldots,K$:

$$
\hat{y}_{lT} (c_k) = \hat{A}'_T \hat{m}_T (c_k), k=1,\ldots, K,
$$

\noindent where $\hat{A}'_T$ is a pseudo-inverse of $\hat{A}_T.$\vspace{1em}

\textbf{Step 4~:} Apply the maximum likelihood estimation based on the joint distribution of $y_l (c_k), k=1,\ldots, K,$ with the pseudo-observations $\hat{y}_{lT} (c_k), k=1,\ldots, K,$ instead of the unobserved $y_l (c_k), k=1,\ldots,K$, to get estimates $\hat{\Theta}_T, \hat{S}_T$ of the long run parameters in (1.6).\vspace{1em}

\textbf{Step 5~:} This approach can be completed by a residual plot that compares the observed values $y$ to fitted values $\hat{y}$  computed on another grid $\gamma_1, \ldots, \gamma_{K^*},$ i.e. the values $y(\gamma_k T)$ to the values~:

$$
\begin{array}{lcl}
\hat{y}_T (\gamma_k T) & = & \hat{\Phi}_T^{(\gamma_k - \gamma_{k-1} ) T} \hat{y}_s (\gamma_{k-1} T) - \hat{y}_{lT} (\gamma_k)\\ \\
&=& \hat{\Phi}_T^{(\gamma_k - \gamma_{k-1})T} [y (\gamma_{k-1} T) - \hat{y}_{lT} (\gamma_{k-1})] + \hat{y}_{lT} (\gamma_k).
\end{array}
$$
\vspace{1em}

Since the emphasis of this paper is on structural dynamic modelling rather than the estimation in a narrow sense, a more detailed discussion of the asymptotic theory addressing the first step estimation errors on $y_l (c_k)$ and of the appropriate choices of $H_T$ and the size $K_T$ of a regular grid on $[0,1]$,  when $T$ tends to infinity is out of the scope of this paper. This would require a functional limit theorem for triangular arrays  [see Gourieroux et al. (2022)].

The distinct rates of convergence of the consistent estimators mentioned in the subsection above, and the impossibility theorem which implies that $\hat{\Theta}_T$ is inconsistent, complicate the efficiency analysis \footnote{The block decomposition introduced in the proposed estimation approach is a time series analogue of the granularity theory for panel data with a dynamic common factor [Gagliardini, Gourieroux (2014)]. We refer to this literature on panel data for the difficulty in defining the notion of efficient estimation when the rates of convergence of estimators are different.}. It follows from the impossibility theorem that when $T \rightarrow \infty$ the accuracy of the consistent approximations of the long run component values $\hat{y}_{lT} (c_k), k=1, \ldots, K_T$, that converge at a speed of order $1/\sqrt{H_T},$ can be disregarded and the accuracy of $\hat{\Theta}_T, \hat{S}_T$, can be negligible when $K_T/H_T \rightarrow 0$. Then the finite sample confidence intervals (CI) of any known scalar nonlinear transformation of $\hat{\Theta}_T, \hat{S}_T$ can be derived by the test inversion [see Kendall, Stuart (1967), Chap. 20, for a description of this technique, Stock (1991), Section 2.3, or Pesavento, Rossi (2006), for an application in a univariate standard local-to-unity model, Carpenter (1999), Dufour (2006) for nuisance parameters and bootstrap extensions].

The estimation steps 1-4 can be related to the literature on mixed frequency data because of the apparent resemblance of combining data measured on one time scale with averaged data measured on a time scale with larger units [see e.g. Deistler et al. (2016), Ghysels (2016)]. Our approach differs from this literature in the following respects:

i) The observations at mixed frequencies (MF) are generated by the same time series, while we consider two different latent processes. In our approach, the observations at different frequencies are used as a tool for identifying the short and long run dynamics, rather than to interpolate the missing observations.

ii) The MF literature considers fixed frequencies, i.e. two fixed time scales. In our framework, we consider an ultra long time scale with a unit that tends to infinity with the number of observations.
Therefore, the impossibility theorem and its consequences for the long run prediction intervals do not concern the MF literature.

iv) Similarly to the MF literature, our estimation approach is based on the Yule-Walker equations at different frequencies to disentangle the short and long run parameters [see e.g. Chen, Zadrozny (1998)].

\section{Long Run Predictions}
\setcounter{equation}{0}\def\theequation{4.\arabic{equation}}

Let us now consider the long run predictions \footnote{The prediction problem is different from the problem of testing the unit root hypothesis and comparing the power functions of the alternative test procedures.}. Our objective is to find the intervals for predictions at large leads as a fixed function of the sample size with a good coverage rate. We assume that the underlying parametric model is well specified.

\subsection{Information Sets}

Corollary 4 implying that the long and short run components can be detected and identified asymptotically simplifies the analysis of predictions, which depend on different information sets. The predictions are conditioned on the current and lagged values of the observed process $\underline{y_t} = (y_t, y_{t-1},...)$. As only  shocks to the ULR component are needed to analyse the systematic long run risk \footnote{Such a practice is facilitated in our framework since the SR and ULR components are independent. This independence assumption and the two time scales provide the structural aspect to the VAR allowing for separate analysis of shocks on the SR and ULR components, respectively.},  it is preferable to work with the information set $(\underline{y_s (t)}, \underline{y_l (t/T)}).$ From Corollary 4  these two information sets are equivalent when $T$ is large.

This result is easy to interpret in the context of the Wold/Volterra decomposition. Let
$(F_t)$ denote the sequence of $\sigma$-algebras generated by process $(y_t)$. The purely predictable component $Z$ is measurable with respect to the  $\sigma$-algebra $F_{-\infty} = \cap_t F_t$ and is stochastic if $F_{-\infty}$ does not degenerate to $({\O}, \mathcal{Y})$. Thus this component belongs  to the  $\sigma$-algebra generated by process $(y_t)$. Even if the observations are available only since an initial date $t=0$ rather than since $- \infty$, the purely predictable component $Z$  can  be recovered asymptotically, if $T \rightarrow \infty$

\subsection{Theoretical Predictions}

Let us consider the predictive distribution $l(y (T+H_T)|\underline{y_{(T)}})$ of $y$ at horizon $H_T$, that is, the conditional distribution of $y (T+H_T)$ given $\underline{y_{(T)}}$. Given the equivalence of the two information sets, the independence between the short and long run components and the Markov assumption on these components, the predictive distribution is:

\begin{equation}
  l (y(T+H_T) | \underline{y(T)}) \approx l_s(y_s (T+H_T)|y_s(T)) * l_l[A y_{l,T} (T+H_T) | y_{l,T} (T)],
\end{equation}

\noindent where $*$ denotes the convolution $l_s, l_l$ the (conditional) distributions of SR and ULR components respectively.

\noindent Two cases can be distinguished:\vspace{1em}

i) For short run predictions, with $H_T/T\rightarrow 0$, when $T \rightarrow \infty$, we have~:

\begin{eqnarray}
  l(y (T+H_T)| \underline{y(T)}) & \simeq & l_s(y_s (T+H_T) | y_s(T)) * \delta_{A y_{l,T}(t)}. \\
                              & = & l_s(y (T+H_T) - A y_{l,T}(T) | y_s (T)),
\end{eqnarray}

\noindent where $\delta_{A y_{l,T} (T)}$ denotes the point mass at $A y_{l,T} (T)$.\vspace{1em}

ii) For long run predictions, with $H_T/T \rightarrow \gamma, \gamma >0$, when $T\rightarrow \infty$, by the ergodicity of the short run component and the fact that it has zero mean~:

\begin{eqnarray}
  l(y(T+H_T) | \underline{y(T)}) & \simeq & l_l [A y_{l,T} (T+H_T) | y_{l,T} (T)] \\
                                & = & l_l [A y_l (1+\gamma) | y_l (1)].
\end{eqnarray}

\noindent Hence, at short horizons, only the short run dynamics and short run component matter, up to a known drift. At long horizons, only the long run dynamics and long run component matter.

\subsection{Sample-Based Predictions}

The theoretical predictive distributions derived in Section 4.2 depend on the unobserved factor values and on the unknown short and long run parameters. In practice, they need to be replaced by the estimated predictive distributions.  This replacement is not innocuous, even for large T, because of the impossibility theorem implying the impossibility to approximate $\Theta$ consistently. There is a long run estimation risk that has to be accounted for in the prediction interval computation. In applied research, it would concern the computation of the Value-at-Risk (VaR),  the required capital (in a financial application of VaR), or deriving the impulse response functions [compare to Gospodinov (2004), Pesavento, Rossi (2006)].

In this section we focus on a combination $X_{T+H_T} = a' Y(T+H_T)$ of variables to be predicted at large horizons $H_T = cT$ and on one-sided prediction intervals. The  one-sided prediction intervals are essential for the computation of the VaR and required capital at horizon $H_T$ under the current financial regulation. The approach below is easily extended to two sided prediction intervals. We denote the set of all model parameters by $\beta$.

\subsubsection{The standard approach}

Before we describe the proposed method, let us recall the standard prediction interval which can be computed when a consistent estimator $\hat{\beta}_T$ of $\beta$ is available.

i) For a given risk level $\alpha$, a theoretical conditional quantile $q$ satisfies the following equality:

\begin{equation}
  P_\beta [X_{T+H_T} < q (1-\alpha, \beta, \underline{Y(T)})|\underline{Y(T)}] = 1-\alpha, \forall \beta.
\end{equation}

\noindent In our Markov framework, it is asymptotically equivalent to~:

$$
  P_\beta [a' Y_s (T+cT) + a' Y_l (T+cT )< q(1-\alpha;\beta, Y_s (T), Y_l(T)) | Y_s(T), Y_l (T)] = 1-\alpha, \forall \beta,
$$

\noindent or, by the independence between the SR and ULR components and the ergodicity of the SR component~:

\begin{equation}
  P_\beta [a' Y_s (T+cT)+ a' Y_l (T+cT) < q (1-\alpha;\beta, Y_l (T))|Y_l (T)] = 1-\alpha, \forall \beta.
\end{equation}

\noindent The only parameters appearing in this expression are the long run parameters $\Theta, S$, and the parameters of the stationary distribution of the short run component (since processes $Y_s$ and $Y_l$ are independent and $Y_s (T)$ no longer appears in the conditioning set).\vspace{1em}

ii) Since the conditional quantile depends on the unknown parameter $\beta$ and on the unobserved long run component, it is common to replace them by their consistent approximations in the expression of the conditional quantile, that is $q(1-\alpha, \hat{\beta}_T, \hat{Y}_l (T)),$ and to introduce the associated estimated prediction interval. Then for large $T$, we have~:

\begin{equation}
  P_\beta [a' X_{T+H_T} < q(1-\alpha; \hat{\beta}_T, \hat{Y}_l (T)) | Y_l (T)] \simeq 1-\alpha, \forall \beta.
\end{equation}

iii) Even if a consistent estimator $\hat{\beta}_T$ is available, in finite sample the estimation risk can exist. It is commonly evaluated by a bootstrap equivalent of a Monte-Carlo experiment based on simulated trajectories with the true parameter value replaced by its estimate $\hat{\beta}_T$ [see Kilian (1999)].

\subsubsection{The approach adjusted for finite sample}

As shown in Section 3.3,  parameter $\Theta$ cannot be consistently estimated, which follows from the impossibility theorem. This implies that the standard approach mentioned above, including possibly the  bootstrap procedure, is no longer valid. Yet, in Section 3.4 we have shown that when  $K_T/H_T \rightarrow 0,$ confidence sets for $\Theta$ , which are valid in finite sample, can be obtained. Then, the remaining parameters, for which consistent estimators are available,  can be replaced by their estimates in the prediction interval formula without entailing significant errors at large $T$.

Let  $I_T (1-\alpha_1; \underline{Y(T)})$ denote a confidence set for parameter $\beta$ at confidence level $1-\alpha_1$. It satisfies~:

\begin{equation}
  P_\beta [\beta \in I_T (1-\alpha_1 ; \underline{Y(T)})] = 1-\alpha_1, \forall \beta.
\end{equation}

\noindent Then, a one-sided sample-based prediction interval at confidence level $1-\alpha$ is obtained by fixing the upper bound of the interval as~:

\begin{equation}
  Q (\alpha, \alpha_1; \underline{Y(T)}) = \max_{\beta \in I_T (1-\alpha_1;\underline{Y_T})} q (1-\alpha + \alpha_1, \beta, \hat{Y}_l (T)),
\end{equation}

\noindent for any choice of $\alpha_1 \in [0, \alpha]$. Indeed we have~:

\begin{equation}
  P_\beta [a' X_{T+H_T} < Q (\alpha, \alpha_1; \underline{Y(T)}) | \underline{Y(T)}] \geq 1-\alpha, \forall \beta, \alpha_1,
\end{equation}

\noindent by applying the Bonferroni inequality.

This sample-based prediction interval depends on the confidence set chosen for the estimation of $\beta$ and  in particular on the level $\alpha_1$. Then, we can also optimize it with respect to $\alpha_1$. In terms of financial regulation, this corresponds to searching for the minimum required capital compatible with level $1-\alpha$. Thus, the one-sided sample-based prediction interval has an upper bound defined by a min-max procedure~:

\begin{equation}
Q^* (\alpha; \underline{Y(T)}) = \min_{0<\alpha_1<\alpha} \max_{\beta \in I_T (1-\alpha_1; \underline{Y_T})} q (1-\alpha+\alpha_1; \beta, \hat{Y}_l(T)).
\end{equation}

\noindent Let $\alpha^*_{1T}, \beta^*_T$ denote the solutions of the min-max optimization (4.12). These solutions depend on $\underline{Y(T)}$ and are fixed endogenously.

The consequences of the additional estimation risk, as compared with the standard computation of Section 4.3.1 are examined below. We have~:

\begin{eqnarray}
  Q^* (\alpha; \underline{Y(T)}) & = & q(1-\alpha;\hat{\beta}_T, \hat{Y}_l (T)) + \{ q(1-\alpha+\alpha^*_{1T};\beta^*_T, \hat{Y}_l (T)) - q(1-\alpha; \hat{\beta}^*_T, \hat{Y}_l(T))\} \nonumber \\
  &=&q (1-\alpha;\hat{\beta}_T,\hat{Y}_l(T)) + \{ q(1-\alpha ; \beta^*_T, \hat{Y}_l(T))- q (1-\alpha;\hat{\beta}_T, \hat{Y}_l(T))\} \nonumber \\
  &+& \{ q(1-\alpha+\alpha^*_{1T}; \beta^*_T,\hat{Y}_l(T)) - q(1-\alpha; \beta^*_T, \hat{Y}_l(T))\}.
\end{eqnarray}

\noindent Formula (4.13) has three terms with the following interpretations:

$\bullet$ the first term is the estimated quantile, that is, the estimated theoretical measure of risk.

$\bullet$ the next two terms represent the estimation risk. The first among them is due to the replacement of approximation $\hat{\beta}_T$ by an extreme $\beta^*_T$ of the confidence set. In the univariate case illustrated below, it is either a lower or upper bound of the confidence interval. The second term is due to adjusting of confidence levels for the lack of accuracy. It increases the confidence interval from $1-\alpha$ to $1-\alpha+\alpha^*_{1T},$ making it more conservative in the light of the definition of a theoretical Value-at Risk (VaR) [Robinson (1977)].

\noindent Let us now illustrate the above approach  in a univariate framework.\vspace{1em}

\textbf{Example  1:} In the univariate model (2.6)-(2.8) the impossibility theorem concerns the scalar parameter $\theta$. All other parameters $\eta, \varphi, s$, as well as the value of the underlying long run component at $T$  can be replaced by consistent estimators and approximations. Therefore, we focus on the theoretical quantile as a function of parameter $\theta$. It is easy to check that for predicting $y(T+H_T)$ at time $T$ and a large horizon $H_T = \gamma T$, the corresponding quantile is:

$$
q(\theta, \alpha) = \exp (-\theta \gamma) y_l (T) + \Psi^{-1} (1-\alpha) \sqrt{\eta^2 + \Frac{s^2}{2\theta} [1-\exp (-2 \theta \gamma)]},
$$

\noindent where $\Psi$ denotes the cumulative distribution function of the standard normal.

Let us now consider a confidence interval for $\theta$ at level $1-\alpha_1$~:

$$
I_T [1-\alpha_1; \underline{Y(T)}] = [\theta_{L,T} (\alpha_1), \theta_{U,T} (\alpha_1)], \; \mbox{say}.
$$

\noindent Since function $q(\theta; \alpha)$ is not necessarily increasing or decreasing in $\theta$, we get~:

$$
Q (\alpha, \alpha_1; \underline{Y(T)}) = \max [q(\theta_{L,T} (\alpha_1), \alpha - \alpha_1), q(\theta_{U,T} (\alpha_1), \alpha - \alpha_1)].
$$

\noindent In the last optimization with respect to $\alpha_1, 0 < \alpha_1 < \alpha$, the optimum will not be reached in practice for $\alpha_1 = 0$, or $\alpha_1 = \alpha$. Indeed for $\alpha_1 = 0$, the confidence interval is usually of infinite length, whereas for $\alpha_1 = \alpha$, that is  $\alpha - \alpha_1 = 0,$ $q(\theta, \alpha - \alpha_1)$ is infinite.

\subsubsection{Bayesian approach and the prudential principle}

The approach developed in Section 4.3.2 assumes a well-specified parametric dynamics for the SR and ULR components. The same assumption underlies the derivation of the one-sided prediction intervals in a Bayesian framework [see e.g. Muller, Watson (2016), (2020), Schorfheide, Song, Yaron (2018) for such approaches). Let us introduce a prior $\pi(\beta)$. Then a Bayesian one-sided prediction interval at level $1-\alpha$ can be defined by a quantile $Q^B (\alpha;\underline{Y_T})$ such that~:

\begin{equation}
  P [a' X_{T+H} < Q^B (\alpha; \underline{Y(T)}) | \underline{Y(T)}] = 1-\alpha,
\end{equation}

\noindent where the conditional probability includes the integration with respect to the "stochastic" parameter $\beta$. More precisely, we get~:

\begin{equation}
  \Int P_\beta [a' X_{T+H} < Q^B (\alpha;\underline{Y(T)})|\underline{Y(T)}] \pi (\beta |\underline{Y(T)}) d \beta = 1-\alpha,
\end{equation}

\noindent where $\pi (\beta |\underline{Y(T)})$ denotes the posterior distribution of $\beta$. As already noted in the literature [see [Robinson (1977), Muller, Norets (2016), Muller, Watson (2020), Section 7.4, for confidence sets versus credible sets], this condition does not ensure that the condition~:

$$
P_\beta [a' X_{T+H} < Q^B (\alpha;\underline{Y(T)}) | \underline{Y(T)}] \geq 1-\alpha, \forall \beta,
$$

\noindent holds uniformly in $\beta$, which is a common characteristic of a prediction interval. This coverage condition is only satisfied in average for a specific posterior (i.e. indirectly for a specific prior). In particular, for financial regulation, {\bf this Bayesian practice is not prudential, as in general it implies an undervaluation of required capital and is very sensitive to the choice of the prior} \footnote{Except in the special case when $P_\beta [\alpha' X_{T+H} < Q^B (\alpha; \underline{Y(T)} | \underline{Y(T)}]$ is independent of $\beta$. In a simple framework, this can arise if the statistic used to construct the prediction interval has some invariance property with respect to the nonidentifiable parameters. But such invariant procedures are difficult to construct for a multivariate $\Theta$.}.  As a consequence, it is not permitted by the current prudential financial supervision.

\section{Illustration}

To illustrate the properties of the SVAR model with a stationary ULR component, we perform a Monte-Carlo study to show the behaviour of various summary statistics. Next we apply the proposed methodology to a set of macroeconomic time series.

\subsection{A Monte-Carlo study}

\subsubsection{The model}

We consider a bivariate model $h=2$ with one long run component $L=1.$ The short run parameters are~:

$$
\Phi = \left(
 \begin{array}{cc}
 0.3 & 0 \\
 0 & 0.7
 \end{array}\right),
 \Omega^{1/2} = \left( \begin{array}{cc} 1 & 1 \\ 0 & 2\end{array}\right),
$$

\noindent with a triangular $\Omega^{1/2}$ matrix to get directional short run causality between these two components. The matrix of "factor sensitivities" is set to~: $A=(1,1)'$ and \;the long run parameters for the univariate long run component to $\theta$ such that $\exp (-10\; \theta ) = 0.4$ and $s=1.$

The simulated series are of length $T=7200,$ which corresponds to about 20 years of 360 days, if the basic time unit is one day. We fix the long run grid to $c_1 =1/20, c_2 = 2/20, \ldots, c_{20} = 20/20$ corresponding to multiples of years, and the length of the short run episode to $H_T = 60$, which is equivalent to two months.\vspace{0.5em}

\noindent Figure 1 provides a simulated path of process $y(t) = (y_{1t}, y_{2t})'$. For each series the effect of the stationary ULR component is masked by the variation of the short run component. We also observe rather parallel smooth waves created by the single latent ULR factor.\vspace{1em}

[Figure 1~: Simulated Trajectory of $y(t)$]\vspace{1em}

\noindent Figure 2 displays the evolutions of the averages $ \hat{m}_1 (c_k), \hat{m}_2 (c_k), k=1,\ldots, 20$ obtained with a bandwidth of 60.\vspace{1em}

[Figure 2~: Trajectory of $\hat{m} (c_k)$]\vspace{1em}

\noindent We observe parallel evolutions of these averages revealing an underlying single long run component.

\subsubsection{The sample autocorrelation functions}

Let us now consider the different types of ACF introduced in Section 3. These are the standard joint ACF computed from $y_t, t=1,\ldots,T,$ up to lag $H=1000$ (Figure 3) and the standard joint ACF computed from $y_t, t=1, \ldots, H_T = 60$, up to lag $H=15$ (Figure 4). The main difference between them is how the autocorrelation is demeaned, i.e. globally in Figure 3, and locally in Figure 4.

\noindent Figure 5 provides the joint ACF computed from 20 observations of the averages $\hat{m} (c_k), k=1,\ldots, K=20$, up to lag 10 of the long run time unit. It is given for bandwidths of 60 and 80, respectively.\vspace{1em}

These figures show the importance of considering jointly the different ACF's. The standard one in Figure 3 reveals the persistence, that is, a slow decay of the auto- and cross-correlations. However, this persistence effect conceals the important features concerning the short and long run components, respectively. By diminishing the length of the time series in Figure 4, and then adjusting for the mean, we reveal the dynamics of the short run component. Next by averaging and changing the time scale in Figure 5, we reveal the dynamics of the long run component.

\medskip

[Figure 3~: The Standard ACF]\vspace{1em}

[Figure 4~: The Local ACF ]\vspace{1em}

[Figure 5~:  ACF of Averages]\vspace{1em}

\subsubsection{Finite sample properties of the ML estimator of $\theta$.}
\setcounter{equation}{0}\def\theequation{5.\arabic{equation}}

Let us  also provide information on the lack of consistency and finite sample distribution of the ML estimator of $\theta$ computed from observations on process $y_l$ over a grid $c_k = k/K$, of $[0,1].$

\noindent We consider $K=25$ used in the application of Section 5.2, and the Gaussian AR(1) model~:

\begin{equation}
  \tilde{y} (k) = \exp (-\theta/K) \tilde{y} (k-1) + \sqrt{1-\exp (-2\theta /K)} \tilde{\varepsilon} (k), k=1, \ldots, K,
\end{equation}

\noindent where the $\tilde{\varepsilon} (k)'s$ are IIN(0,1). Without loss of generality the  scale parameter is set equal to 1, since parameter $\theta$ is scale invariant. Parameter $\theta$ is estimated by the maximum likelihood conditional on $y (0)$ where $y (0)$ is drawn in $N(0,1)$.

For each level $\alpha, \alpha = 0.05, 0.1, 0.5, 0.9. 0.95$, we derive the one sided $\alpha$- percentiles interval for $\hat{\rho}_K = \exp (-\hat{\theta}/K)$ as a function of $\rho=\exp (-\theta/K)$. This provides belts that can be inverted to get the $1-\alpha$ confidence intervals for $\rho$ from an estimate $\hat{\rho}_K$.
The sudden change of slope in the 0.95 quantile occurs when $\rho$ is too large, It is due to the finite sample properties of the estimator close to a unit root [see Stock (1991), Figure 2, in the pure unit root case].

By comparing the estimate with the 50\% quantile, we get information on the skewness of the finite sample distribution of the estimator of $\rho$.
\medskip

\nin [Insert Figure 6: Confidence Belts]

\subsection{Application to Economic Time Series}

\subsubsection{The series}

As an empirical exercise and to facilitate the comparison with the recent literature [Muller, Watson (2016), (2020)], we consider the same five quarterly US macro-series as those examined by Muller, Watson in their papers. The series obtained from the Federal Reserve Bank of St Louis FRED database are the growth rates of total factor productivity (TFP),  per-capita GDP, consumption, investment and labor compensation. They are not annualized and cover the period 1948 Q2-2019 Q2 prior to  the COVID pandemic. The evolutions of these series are displayed in Figure a.1, which corresponds to Figure 2 in Muller, Watson (2020). We give in Figure a.2 in Appendix 4 the associated ACF and cross-ACF.

The standard tools of time series analysis reveal no persistence, i.e. no remaining "unit root" features in the growth rates. However, in the framework of a model with SR and ULR components, the standard ACF provides no clear information about the ULR. The ACF at short lags reveals only the persistence of the short run component. The autocorrelations at very long lags are close to zero due to the large variance of the SR component relatively to the variance of the underlying ULR component.

Following the standard approach, ARMA models are fitted to the series without taking into account the underlying structure represented as the sum of a SR and an ULR components. In particular, they are estimated by OLS-type estimators that optimize the SR prediction performance at horizon 1, whereas we are interested in the ULR prediction performance. Nevertheless we estimate an ARMA model for each series and then compute the standard prediction intervals at a date corresponding to $T=275$. Consistently with our proposed ULR approach, we choose the ultra-long horizon equal to $H=200$ quarters, i.e. 50 years \footnote{Different ULR horizons corresponding to other fractions of $T$ can also be considered, such as $H=120$, i.e. $30$ years.}. The standard prediction intervals are given in the first column of Table 2.

\subsubsection{Filtering the ULR component}

To filter out the potential ULR component, we partition the time scale measured in quarters into non-overlapping intervals of length 11 quarters. Then, the raw data are averaged over each interval and their average is assigned to the middle date of each interval. We get new series of length 25 on a deformed time scale of 21 quarters. Although this filtering approach follows the same idea as the low frequency approach of Muller, Watson (2016), (2020), it differs from their approach
with respect to the following. Muller, Watson base their analysis on projections on a fixed number of low frequency periodic functions of time, with frequencies such as $2T/j, j=1, \ldots, 14$ [see Muller, Watson (2020), Section 3]. This approach uses all observations to construct the summary statistics. Therefore, Muller and Watson's approach is a global-in-time approach while ours is local-in-time. A drawback of the global-in-time approach is easy to point out in the context of a model with two additive components, which are a ULR component and a white noise as the SR component. It is well-known that the spectrum of white noise process is characterized by positive equal weights at each frequency. Hence, the global-in-time projection approach does not eliminate the effect of the noise, and as a consequence does not focus on the ULR component only. From a purely theoretical perspective, this approach shows a lack of coherency with respect to the Wold representation. While the modelling assumes a regular infinite order moving average process, the projection is intended to detect singular components (which are stochastic combinations of deterministic periodic functions of time). Our approach distinguishes between the regular and singular (i.e. SR and ULR) components and attenuates the effects of the SR component in order to filter out the ULR component.
The ACF and cross-ACF of the series of filtered components are provided in Figure a.3 in Appendix 4. The AR(1) models are estimated from each filtered serie. The values of estimated autoregressive coefficients in the ULR time scale and the calendar time scale measured in quarters are given in Table 1.
\vspace{1em}

\centerline{Table 1~: Autogressive Coefficients for the ULR Component}
\vspace{1em}

\begin{tabular}{|l|c|c|c|c|c|}
  \hline
  Series & tfp & gdp & c & inv & lab. \\ \hline
  ULR Scale & $(-) 0.275$ & $(-) 0.139$ & $(-) 0.025$ & $ 0.546$ & $ 0.160$ \\
  quarter scale & $0.889$ & $0.836$ & $0.714$ & $0.947$ & $0.847$ \\
  \hline
\end{tabular}
\vspace{1em}

Due to time deformation, the small values of autregressive coefficients in the ULR scale becomes much larger and closer to 1 in the calendar time scale measured in quarters.

Next, each estimated AR(1) model is used to predict at the same ULR horizon as described in Section 5.2.1 corresponding to about $H=18$ in the ULR time scale. \footnote{Some filtered ULR series show negative autocorrelation coefficients. They seem to be due to the special cleaning treatment of the series in the database. Since the prediction formulas depend on $\rho^2$ only, the sign has no impact on the prediction intervals.} We provide the plug-in theoretical prediction intervals in the second column of Table 2, and the intervals adjusted for estimation risk (see Section 4.3.2) in the third column. The differences between columns 2 and 3 show the importance of taking into account the estimation risk in ULR predictions due to the significant increase of $10\%$ to $200\%$ in the length of the prediction formula in the third column.

These results can be compared with the prediction intervals computed from the series of raw data. The prediction intervals based on raw data are shorter than the adjusted filtered ULR prediction intervals. In addition, they are biased, since they assume a constant mean instead of a slowly varying mean.
Our results cannot be directly compared to the prediction intervals derived for the same series in Muller, Watson as they derived prediction intervals for the change averages between $T$ and $T+H$ at $90\%$. Because of the averaging and using critical values associated with lower confidence levels, Muller, Watson report shorter prediction intervals.\vspace{1em}

\centerline{Table 2~: Prediction Intervals at $95\%$}
\vspace{1em}
\begin{tabular}{|c|c|c|c|}
  \hline
  Series & Raw Data & Filtered ULR & Filtered ULR \\
         &          & (unadjusted) & (adjusted for\\
         &          &              &  estimation risk) \\ \hline
  tfp & $(0.781, 1.634)$ & $(0.437, 1.436)$ & $(0.108, 1.763)$ \\ \hline
  gdp & $(0.0065, 0.0090)$ & $(0.0054, 0.0081)$ & $(0.0048, 0.0087)$ \\ \hline
  c  & $(0.0069, 0.0089)$ & $(0.0070, 0.0094)$  & $(0.0067, 0.0096)$ \\ \hline
 inv & $(0.0065, 0.0085)$ & $(-0.0017, 0.0113)$ & $(-0.0136, 0.0231)$ \\ \hline
  lab & $(0.0065, 0.0091)$ & $(0.0076, 0.0111)$ & $(0.0069, 0.0119)$ \\
  \hline
\end{tabular}

\bigskip

\section{Concluding Remarks}

The aim of this paper is to introduce stationary long run components for stationary SVAR models and to derive long run prediction intervals for this model. It follows from the impossibility theorem that the long run predictions have to account for a significant long run estimation risk that makes the prediction intervals significantly wider.

The ULR framework is an alternative to other local-to-unity models introduced in the literature, where the long run component leads to  explosive evolutions, with trajectories that either explode in the ultra long run, or take extreme values with high probabilities (see Appendix 1). This questions the validity of long run predictions based on these models for a variety of economic or financial variables, which are not explosive and take values from limited sets.

As shown in this paper, by applying well-chosen block decomposition of  observations, it is possible to separate the statistical inference on short and long run parameters as well as to identify the short and long run components of the process. The main message of this paper is related to the impossibility theorem, and conveys the impossibility to consistently estimate some long run parameters. This implies that estimation risk is significant and has to be taken into account in the long run prediction intervals, as well as in the VaR and required capital computation methods. While the proposed approach is clearly model-based, from a prudential perspective the long run estimation risk should not be artificially underestimated by either using an over-scaled local-to-unity model (see Appendix 1), or a Bayesian prediction approach that does not ensure the uniform coverage of the prediction interval (Section 4.3.3). More generally, prediction methods that appear misleadingly precise because of underestimating the long run risk, need to be avoided.

The analysis presented in the paper can be extended to additive decompositions of short and long run component with non- Gaussian distributions, to avoid the standard identification issues in the second-order analysis of SVAR [see Gourieroux, Monfort, Renne (2017)]. It can also be extended to nonlinear dynamic models with  a non-additive decomposition into the short and long components with independent nonlinear dynamics of those components [see Gourieroux et al. (2021)].

\newpage
\centerline{Figure ~1 : Simulated Trajectory of Bivariate Process $y(t)$}\vspace{1em}
\includegraphics[width=\linewidth]{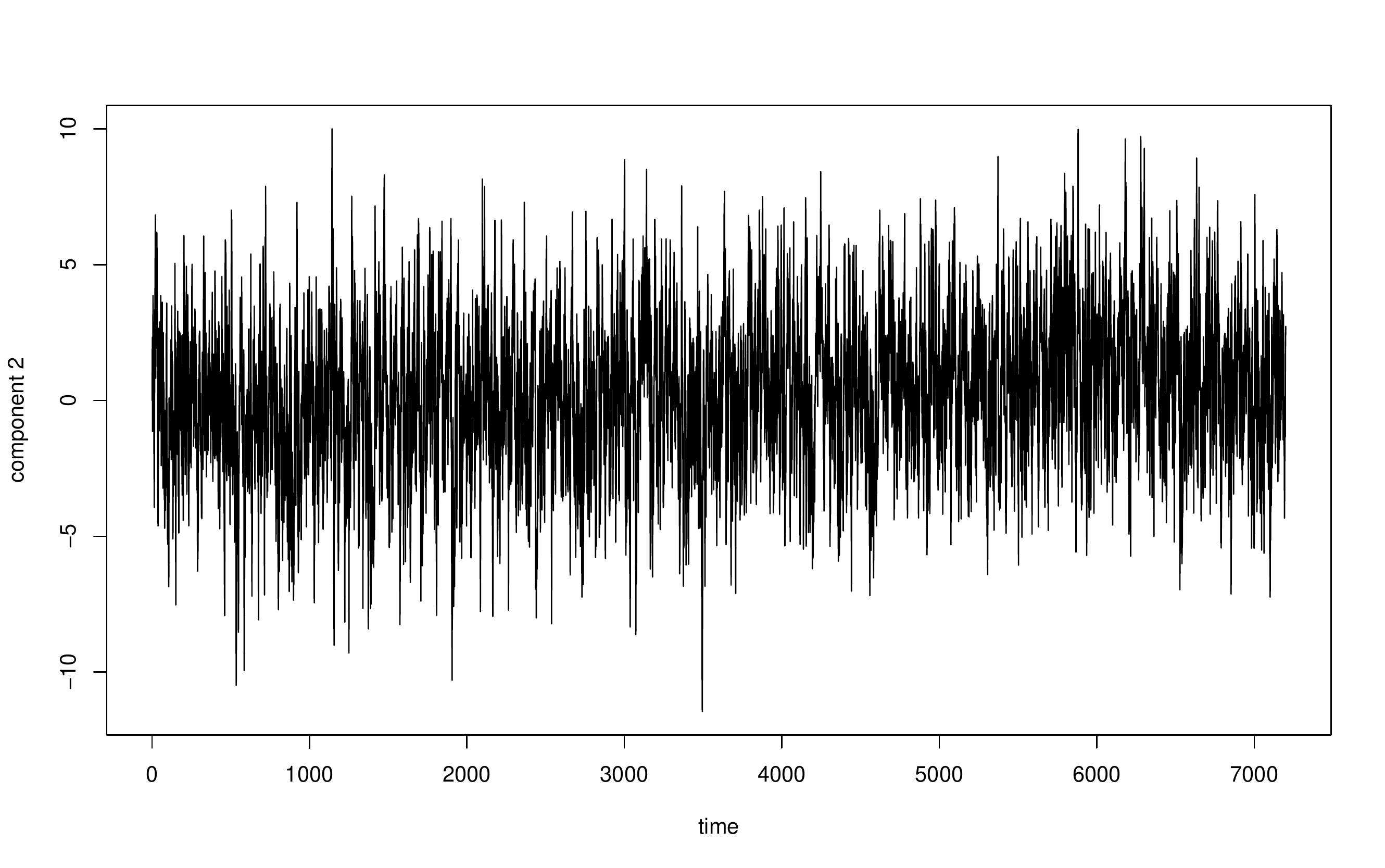}
\includegraphics[width=\linewidth]{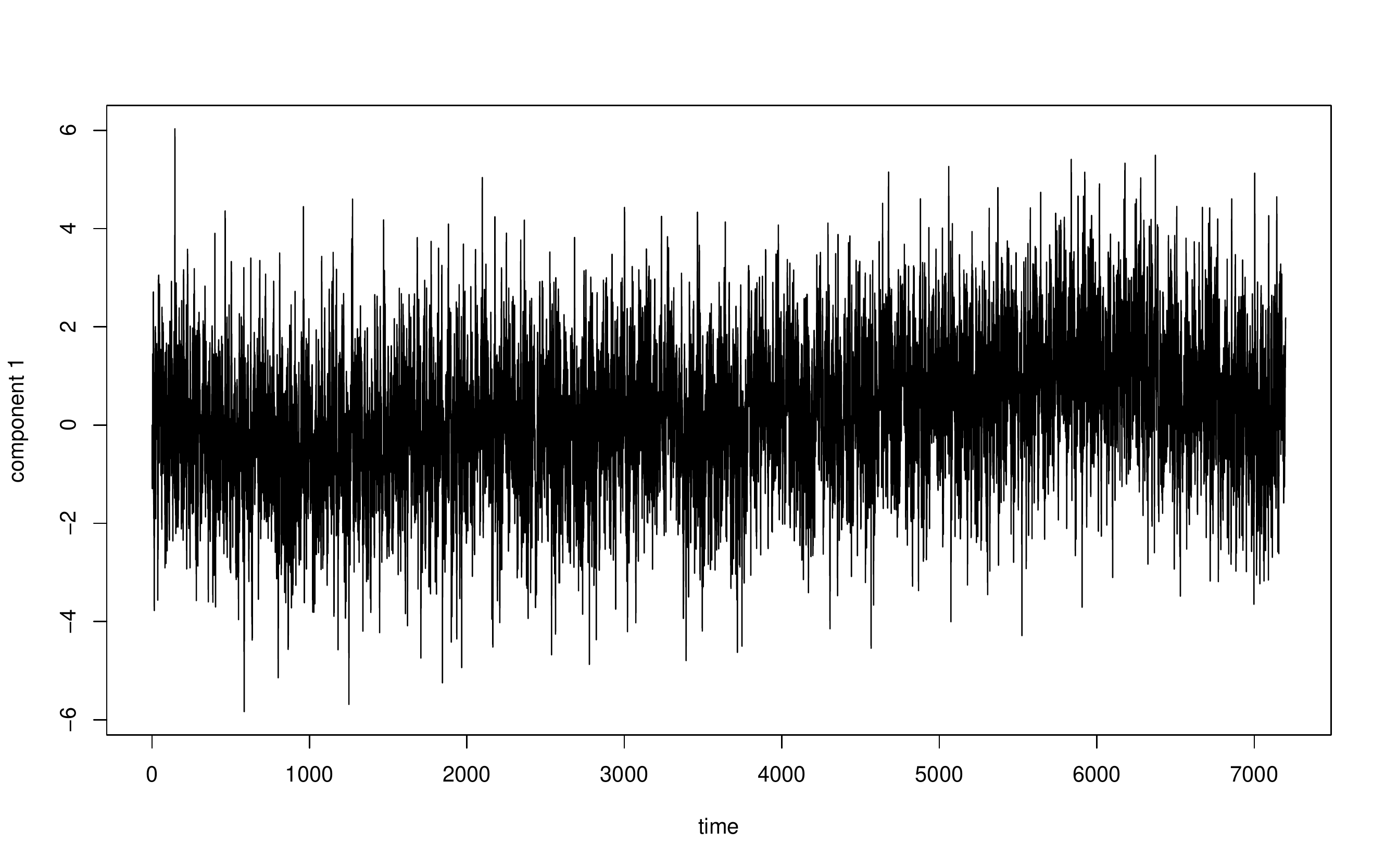}
\newpage
\centerline{Figure ~2 : Trajectory of $\hat{m} (c_k)$}\vspace{1em}
\includegraphics[width=\linewidth]{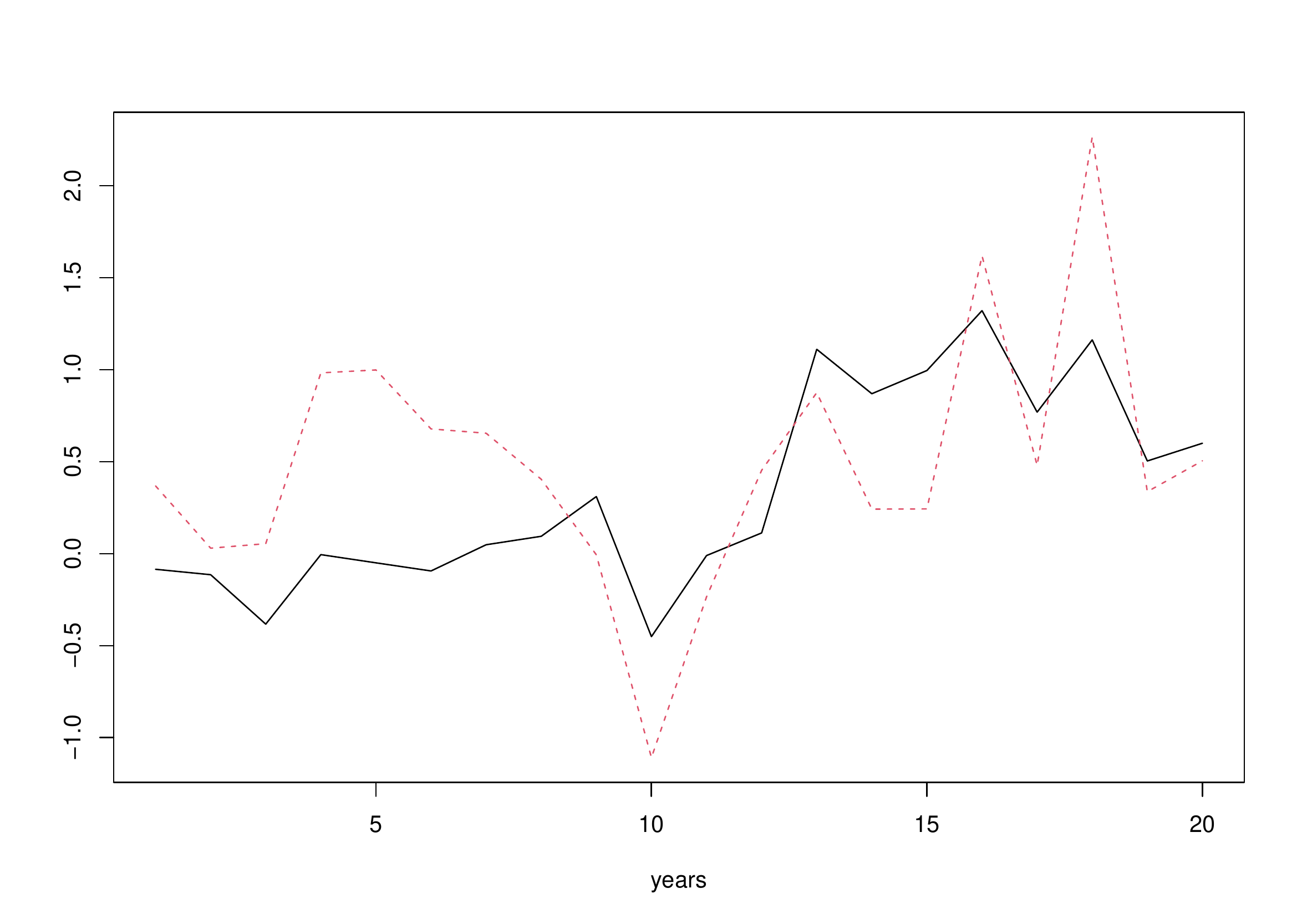}
\newpage

\centerline{Figure ~3 : The Standard ACF}\vspace{1em}
\includegraphics[width=\linewidth]{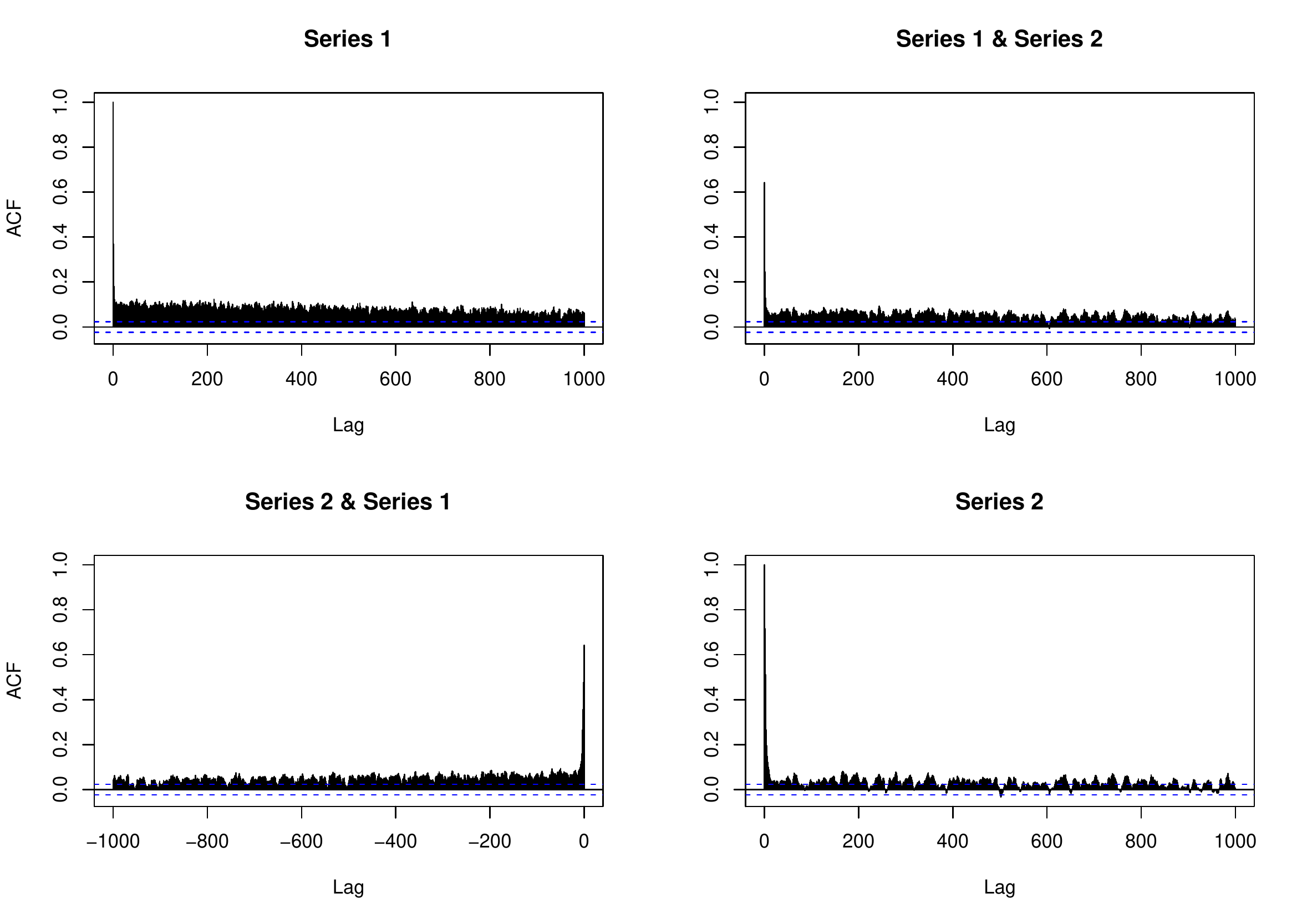}
\newpage
\centerline{Figure ~4 : The Local ACF}\vspace{1em}
\includegraphics[width=\linewidth]{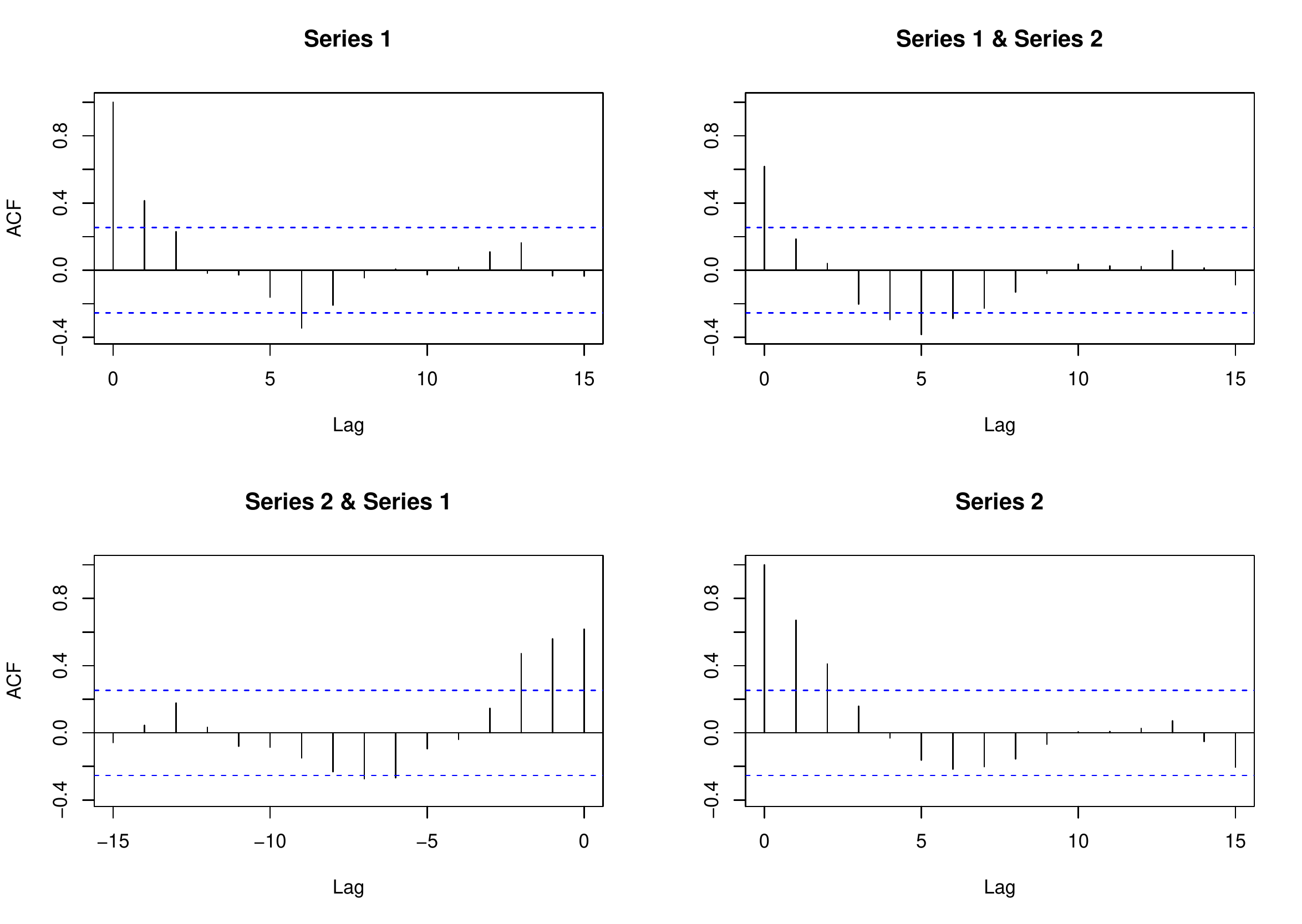}

\newpage
\centerline{Figure ~5 : ACF of Averages}\vspace{1em}
\includegraphics[width=\linewidth]{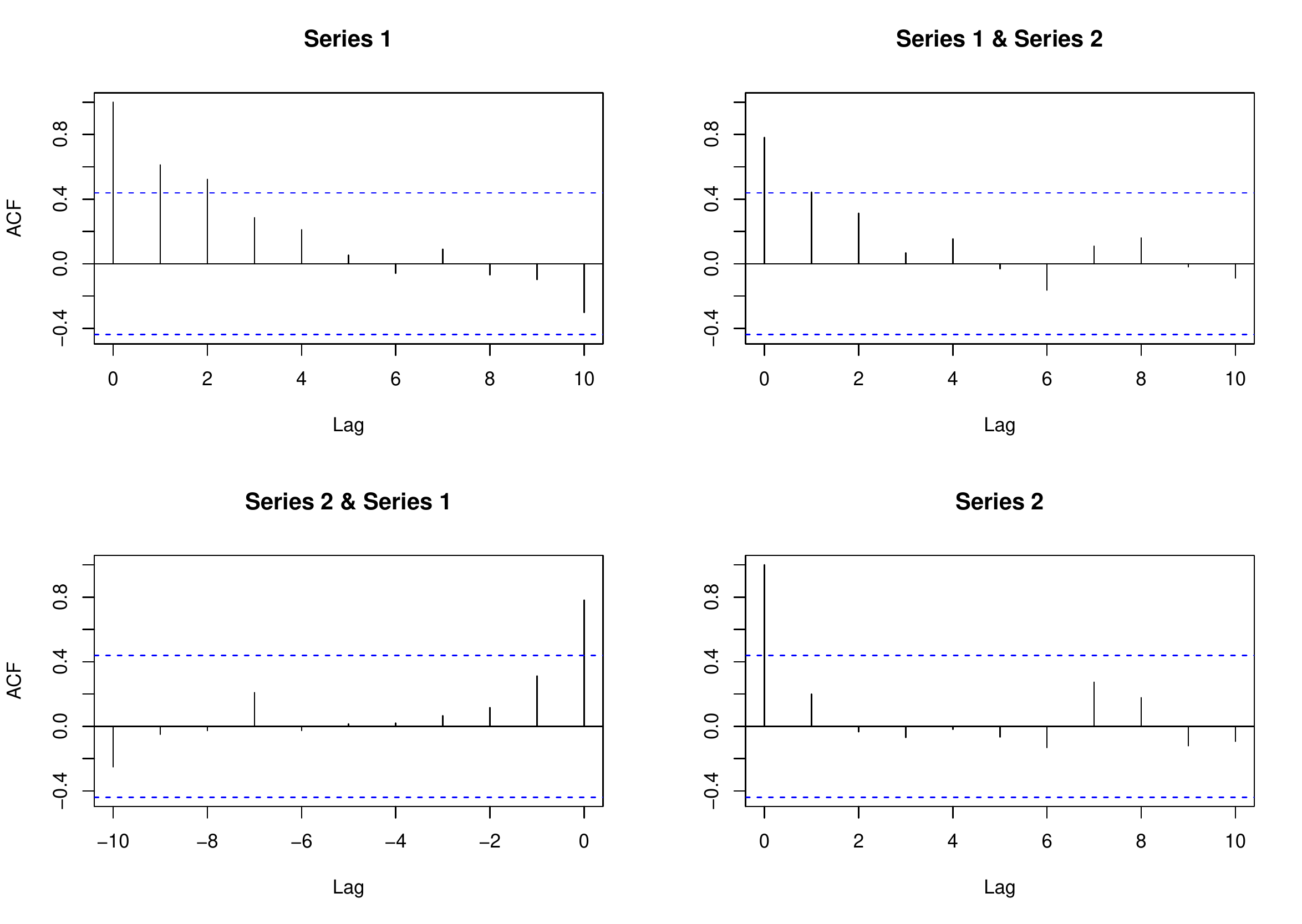}
\centerline{bandwidth = 60}\vspace{1em}

\includegraphics[width=\linewidth]{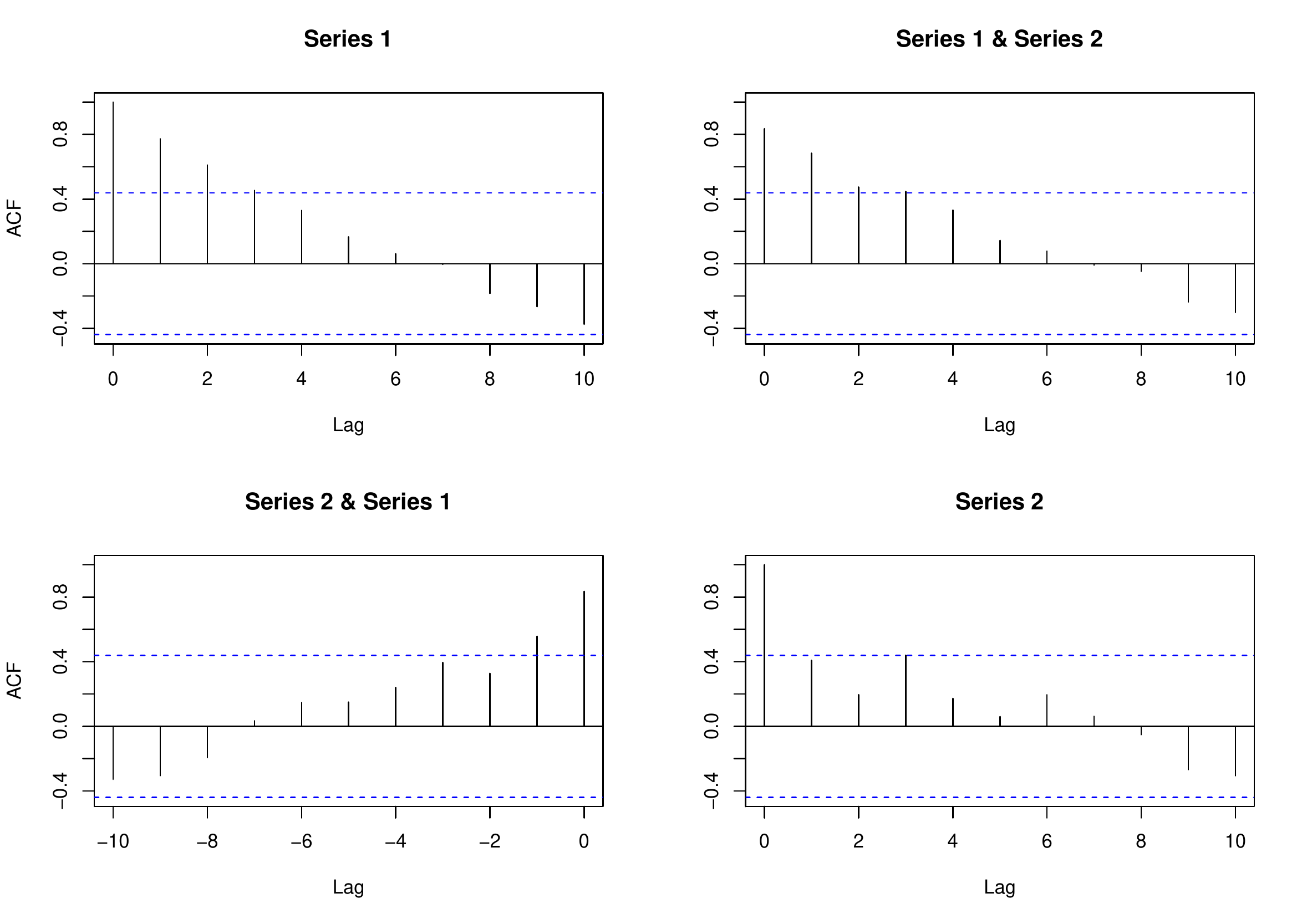}
\centerline{bandwidth= 80}\vspace{1em}

\newpage
\centerline{Figure 6: Quantiles of Estimated $\rho$ at 0.05, 0.1, 0.5, 0.9 and 0.95}\vspace{1em}

\centerline{\includegraphics[width=10cm]{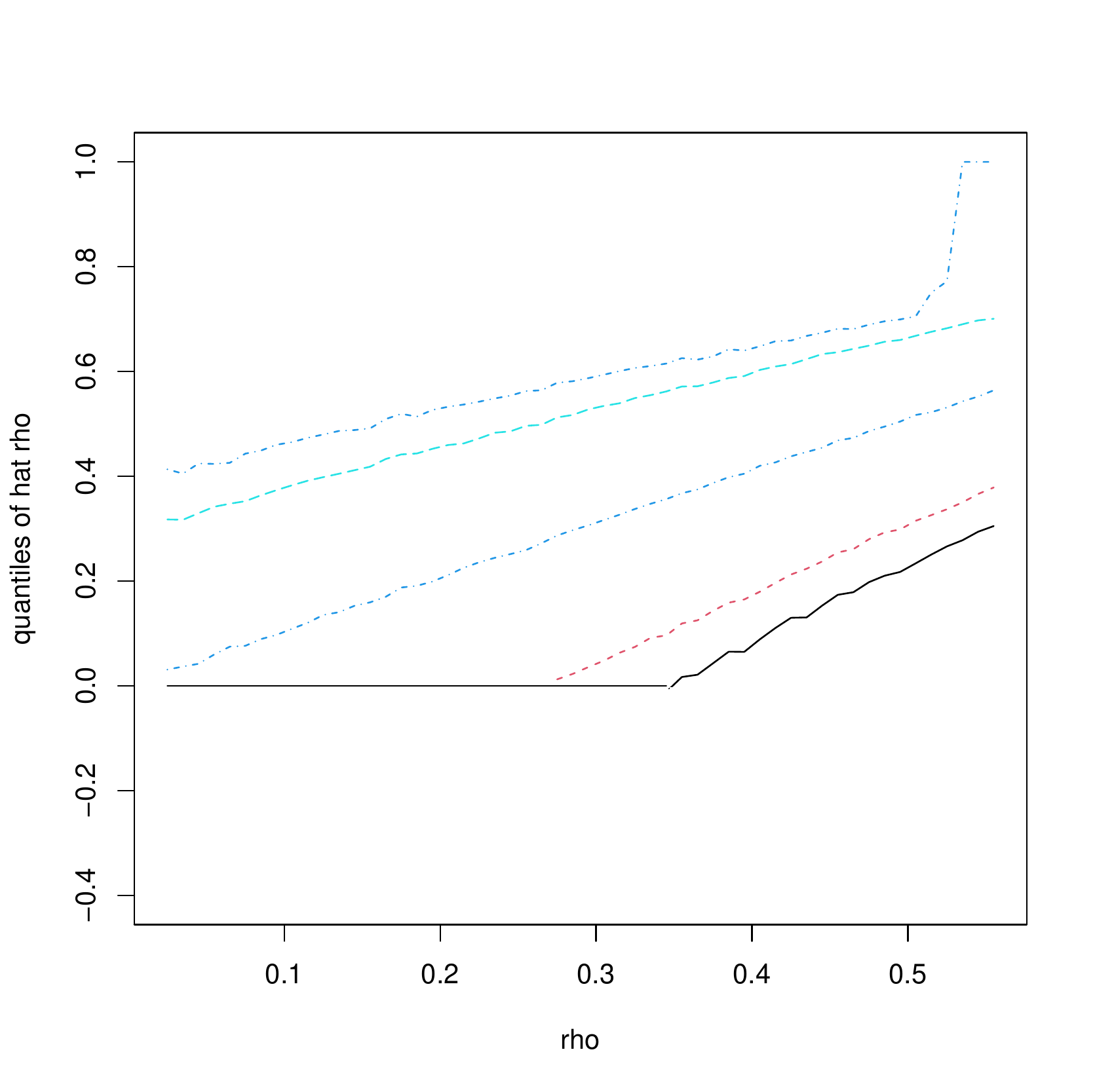} }

\newpage
     \begin{center}
\textbf{Appendix 1}\vspace{1em}

\textbf{Local-To-Unity (LTU) Models}
\end{center}
\vspace{1em}

This appendix  provides a brief survey of the different types of LTU models introduced in the literature to compare their ability of representing the dynamics of time series featuring "persistence" in sample autocorrelations  and a significant long run risk. The condition on the extremes is~: $\sup_t P [|y_t|> A] \equiv \pi (A) < 1$, say. The above condition eliminates explosive trends as well as explosive variances. For expository purpose, we consider the univariate (AR(1)) Gaussian process, where $(\varepsilon_t)$ denotes a Gaussian white noise.

Initially the aim of the LTU models was to approximate a unit root model such that $Corr (y_t, y_{t-1}) = 1.$ Recall that there exists two types of unit root Gaussian AR(1) models:\vspace{1em}

i) \textbf{The random walk}
\setcounter{equation}{0}\def\theequation{a.\arabic{equation}}

$$
y_t = y_{t-1} + \sigma \varepsilon_t, \sigma > 0,
$$

\noindent where $y_0$ is either deterministic, or stochastic, independent of $\varepsilon_t, t > 0$. This process is such that $y_t = y_0 + \Sigma^t_{\tau = 1} \varepsilon_\tau$. It is nonstationary and "explosive" with $\sup_t P (|y_t|>A) = 1.$

Such models have for instance be applied to time varying parameters (mean and log-variances) to analyze the long run GDP growth in Antolin-Diaz et al. (2017).\vspace{1em}

ii) \textbf{The singular process}\vspace{1em}

$$y_t = y_{t-1},$$

\noindent with a stochastic $y_0$. This is a stationary process such that $\sup_t P [|y_t|>A] = P (|y_0 | > A) = \pi (A) <1.$ It has no explosive features.\vspace{1em}

\noindent Let us now consider local-to-unity models, which are written for triangular arrays, $y_T (t)$. Thus the condition on the extremes is expected to be uniform in $T$:

\begin{equation}
  \sup_T \sup_{t \leq T} P_T [|y_T(t)|>A] = \pi (A) <1.
\end{equation}

a) \textbf{Local-to-Unity with 0 initial value} [Chan, Wei (1987), Phillips (1987)]\vspace{1em}

$y_T(t) = \exp (-c/T) y_T(t-1) + \sigma \varepsilon_t$, with $y_T(0) = 0, c>0.$\vspace{1em}

\noindent The distribution of $y_T(t)$ is $N\left(0, \sigma \Frac{1-\exp (-2ct/T)}{1-\exp (-2 c/T)}\right)$.

In particular $y_T(\gamma T) \sim N \left[0, \sigma^2 \Frac{1-\exp (- 2c\gamma)}{1-\exp (2c/T)}\right], \gamma > 0$.

At dates equal to multiples of the number of observations, we see that $P_T [|y_T (\gamma T)| >A] \rightarrow 1$, for $T \rightarrow \infty$. {\bf The condition on the extremes is not satisfied}.\vspace{1em}

b) \textbf{Stationary Local-to-Unity} [Elliott (1999), top of p820, Elliott, Muller (2003)].

$$
y_T(t) = \exp (-c/T) y_T(t-1) + \sigma \varepsilon_t,
$$

\noindent where the initial value $y_T(0)$ is drawn in the unconditional distribution $N (0, \Frac{\sigma^2}{1-\exp (-2c/T)})$.

When $T \rightarrow \infty$, the unconditional distribution of $y_T(t)$ tends to $N(0, \infty)$, and {\bf the condition on the extremes is not satisfied}.\vspace{1em}

c) \textbf{Scaled Local-to-unity with $1/T$ scale effect} [Muller, Watson (2016), footnote 9, Muller, Watson (2020), p11].

$$
\left\{
\begin{array}{lcl}
y_T(t) & = & \exp (-c/T) y_T (t-1) + \Frac{\sigma}{T} \varepsilon_t, \\ \\
y_T(0) & = & 0.
\end{array}
\right.
$$

\noindent In this case, we get~:

$$
y_T (\gamma T) \sim N (0, \Frac{\sigma^2}{T^2} \Frac{1-\exp (-2c \gamma)}{1-\exp (-2c/T)}) \approx N(0,0),
$$

\noindent when $T$ tends to infinity. {\bf The scale effect is too strong leading to a deterministic constant zero at long horizons and an artificial total diversification of the long run risk}.\vspace{1em}

d) \textbf{Scaled Random Walk} [Muller, Watson (2020), p11]

$$
\left\{
\begin{array}{lcl}
y_T(t) & = & y_T (t-1) + \Frac{\sigma}{T} \varepsilon_t, \\ \\
y_T(0) & = & 0.
\end{array}
\right.
$$

\noindent Similarly as for the scaled LTU in c) we have~:

$$
y_T (\gamma T) \sim N (0, \Frac{\sigma^2}{T^2} \gamma T) = N (0, \Frac{\sigma^2 \gamma}{T}) \approx N (0,0),
$$

\noindent when $T$ tends to infinity. {\bf The scale effect is too strong}.\vspace{1em}

e) \textbf{The Ultra Long Run (ULR)}\vspace{1em}

This dynamic model bridges the gap between the standard LTU models a), b) and the scaled LTU in c). It is defined as~:

$$
y_T(t) = \exp (-c/T) y_T(t-1) + \sigma \sqrt{1-\exp (-2c/T)} \varepsilon_t.
$$

\noindent For each fixed $T$, we get a stationary process as in example b), but due to the choice of the variance, the unconditional distribution $N(0,\sigma^2)$ no longer depends on $T$. {\bf Therefore the uniform condition for the extreme becomes}:

$\sup_T \sup_{t \leq T} P_T [|y_T (t)|<A] = P(\sigma \mathcal{U}< A) = \pi(A)<1,$ where $\mathcal{U} \sim N [0,1]$, and is {\bf satisfied}.\vspace{1em}

When $T \rightarrow \infty$, the dynamic is close to the "dampered unit root" process~: $y_T (t) = (1-\Frac{c}{T}) y_T (t-1) + \lambda/\sqrt{T} \varepsilon_t$, considered in Gospodinov (2009), Gospodinov, Maynard, Pesavento (2021).\vspace{1em}

f) \textbf{Other time deformation models}\vspace{1em}

The ULR model can be extended to other time deformations, for instance corresponding to a large time unit of $T^d, d>0$. It is easy to show that the ULR model corresponding to $d=1$ is a limiting case. Indeed, by taking appropriate averages of the observations, we observe asymptotically the underlying OU process on $(0, \infty),$ if $d<1$. Therefore parameter $c$ (in the notation of this appendix, $\theta$ in the notation of the text) can be estimated consistently. In some sense the ULR model is the first prudential model in this class of time deformed autoregressive models.

\newpage
     \begin{center}
\textbf{Appendix 2}\vspace{1em}

\textbf{Asymptotic Behaviour of $\Frac{1}{T-cT} \Sum^T_{t=cT+1} [y_T (t) y'_T (t-cT)]$}
\end{center}
\vspace{1em}

We have~:

$$
\begin{array}{lcl}
 && \Frac{1}{T-cT} \Sum^{T}_{t=cT+1} [y_T (t) y'_T (t-cT)] \\ \\
& =& \Frac{1}{T-cT} \Sum^{T}_{t=cT+1} [y_s (t) y'_s (t-cT)] + A \Frac{1}{T-cT} \Sum^T_{t=cT+1} [y_{lT} (t) y'_{lT} (t-cT)]A' \\ \\
 &+ & \Frac{1}{T-cT} \Sum^T_{t=cT+1} [y_s (t) y'_{lT} (t-cT)]A'
+ \Frac{A}{T-cT} \Sum^T_{t=cT+1} [y_{lT} (t) y'_s (t-cT)].
\end{array}
$$

The first term of the right hand side of the equality tends to zero in probability by the standard Law of Large Numbers for stationary geometrically mixing Gaussian processes.

The second term tends to the corresponding stochastic integral by applying the convergence of Riemann sums [Stroock, Varadhan (1979), Section 11,]

With respect to the discussion on asymptotic behaviour of sample means, the two remaining terms representing the cross-effects of the short and long run components are new. We have to check if they are negligible. This is a consequence of the weak Law of Large Numbers for mixingales [De Jong (1998)].

Since the weak LLN for a triangular array is usually discussed in the univariate process, we consider below the case $n=L=1$ for expository purpose. The approach is easily extended to matrices by applying the same reasoning element par element. Let us consider the fourth term and denote :

$$
X_{Tt} = y_{l,T} (t) y_s (t-cT) = y_l (t/T) y_s (t-cT),
$$

\noindent and $\mathcal{F}_{Tt} = \left( \underline{y_l (t/T)}, \underline{y_s (t-cT)}\right)$ an infinite array of $\sigma$-fields.\vspace{1em}

a) \textbf{$L_1$-mixingale condition}\vspace{1em}

Let us check that $(X_{Tt}, \mathcal{F}_{Tt})$ satisfies the condition of a $L_1-$mixingale array [Andrews (1988), De Jong (1998), Definition 1].

i) By construction $X_{Tt}$ is measurable with respect to the increasing sequence of $\sigma$-fields $\mathcal{F}_{Tt}, t$ varying. Therefore condition (2) in De Jong (1998), definition 1, is trivially satisfied [see Andrews (1988), p460, remark b)].\vspace{1em}

ii) Let us now consider condition (1) in De Jong definition 1. We have~:

$$
\begin{array}{l}
  ||E [X_{Tt} | \mathcal{F}_{T,t-m}] ||_1 \\ \\
  = || E [y_l (t/T)|y_l \left( \Frac{t-m}{T}\right) ] E [y_s (t-cT)|y_s (t-m-cT)]||_1,
\end{array}
$$

\noindent by the independence between the SR and ULR components and their Markov properties. Then,

$$\begin{array}{l}
|| E [X_{Tt} | \mathcal{F}_{T,t-m}]||_1\\ \\
= || \exp (-\theta m/T) y_l (\Frac{t-m}{T}) \varphi^m y_s (t-m-cT)||_1 \\ \\
\leq || y_l (\Frac{t-m}{T}) y_s (t-m-cT)||_1 \varphi^m \equiv K \psi (m),
\end{array}
$$

\noindent where $K$ is a constant by the independence of the SR and ULR components and their stationarity, and where $\psi (m) = \varphi^m$ tends to zero when $m$ tends to infinity. Thus condition (2) is satisfied.\vspace{1em}

b) \textbf{Weak LLN}\vspace{1em}

There exist different versions of weak LLN for $L_1$-mixingales. The standard version in Andrews (1988), Theorem 1, is not applicable in our framework due to a too strong condition of uniform integrability of $X_{Tt}$. However, we can apply De Jong (1995), Theorem 1 [see also De Jong (1998), p 214]. In his notations, the conditions are satisfied with $B_T = 1$ constant. Indeed we have~:

$$
\begin{array}{l}
\lim_{K\rightarrow \infty} \lim \sup_{T \rightarrow \infty} \Frac{1}{T} \Sum^T_{t=1} || X_{Tt} \id_{[X_{Tt}]>K}||_1 \\ \\
= \lim_{K\rightarrow \infty} || X_{Tt} \id_{[X_{Tt} | > K}||_1\; \mbox{by stationarity,}\\ \\
= 0\;, \mbox{since the expectation of}\; X_{Tt}\; \mbox{exists.}
\end{array}
$$

\newpage
     \begin{center}
\textbf{Appendix 3}\vspace{1em}

\textbf{Asymptotic Behaviour of Local Means}
\end{center}
\vspace{1em}

The local mean at date $cT$ is defined by [see eq. (3.4)]~:

$$
\begin{array}{lcl}
\hat{m}_T (c) & = & \Frac{1}{H_T} \sum^{cT+H_T}_{t=cT+1} y_T (t) \\ \\
&=& \Frac{1}{H_T} \sum^{cT+H_T}_{t=cT+1} y_s (t) + \Frac{1}{H_T} \sum^{cT+H_T}_{t=cT+1} A y_l (t/T).
\end{array}
$$

By the independence assumption the two components of the right hand side can be analyzed separately.\vspace{1em}

\textbf{a) Analysis of the SR component}\vspace{1em}

Since the SR process is stationary, the sample average $\hat{m}_{sT} (c) =\Frac{1}{H_T} \sum^{cT+H_T}_{t=cT+1} y_s (t)$ has the same distribution as the average $\Frac{1}{H_T} \sum^{H_T}_{t=1} y_s (t)$. Therefore under the assumption of geometric ergodicity of $y_s$ (that is the modulus of the eigenvalues of $\Phi$ strictly smaller than 1) and the existence of its second-order moments, we have~:\vspace{1em}

i) $\hat{m}_{sT} (c) \rightarrow 0$, if $H_T \rightarrow \infty$, \vspace{1em}

ii) $\sqrt{H_T} \hat{m}_{sT} (c) \rightarrow \Sigma^{1/2}_{\infty} Z_s (c)$,\vspace{1em}

\noindent where $Z_s (c) \sim N(0, Id),$ and $\Sigma_\infty = \sum^{+\infty}_{h=-\infty}\; \mbox{cov}\; [y_s (t), y_s (t-h)].$ \vspace{1em}

iii) Moreover, for different levels $c_k, k=1,\ldots, K$, the limiting $Z_s (c_k), k=1,\ldots, K$ can be chosen independent, if $H_T/T \rightarrow 0$.\vspace{1em}

\textbf{b) Analysis of the second component}\vspace{1em}

Let us consider the difference~:

$$
\Delta_T (c) = \Frac{1}{H_T} \sum^{cT+H_T}_{t=cT+1} y_l (t/T) - y_l (c).
$$

Since $(y_l)$ is a Gaussian process, $\Delta_T (c)$ is also a Gaussian process. We have~:

$$
\begin{array}{lcl}
\sqrt{H_T} \Delta_T (c) & = & \Frac{1}{\sqrt{H_T}} \sum^{cT+H_T}_{t=cT+1} [y_l (t/T) - y_l (c)] \\ \\
                        & = & \Frac{1}{\sqrt{H_T}} \sum^{H_T}_{h=1} [y_l (c+h/T) - y_l (c)].
\end{array}
$$

The Ornstein-Uhlenbeck process has a closed form expression with respect to its value $y_l(c)$ at $c$ and the future values of the Brownian motion as~:

$$
y_l (c+h/T) = \exp (- \Theta h/T) y_l (c) + \exp (-\Theta h/T) \Int^{c+h/T}_c \exp (\Theta u) S d W_u.
$$

Then the asymptotic behaviour of $\sqrt{H_T} \Delta_T (c)$ depends on the independent effects of $y_l (c)$ and the future of the Brownian motion $W_{c+u}$.\vspace{1em}

i) Effect of $y_l (c)$\vspace{1em}

The component of $\Delta_T (c)$ depending on $y_l (c)$ is~:

$$
\begin{array}{lcl}
\Delta_{1T} (c) & = & \Frac{1}{H_T} \sum^{H_T}_{h=1} [\exp (-\Theta h/T) - Id] y_l (c) \\ \\
&\simeq& -\Frac{1}{H_T} \sum^{H_T}_{h=1} [\Theta h/T] y_l (c), \; \mbox{if}\; H_T/T\rightarrow 0, \; \mbox{when}\; T \rightarrow \infty, \\ \\
&\simeq& -\Frac{\Theta}{TH_T} \sum^{H_T}_{h=1} h y_l (c) \\ \\
&\simeq & -\Frac{H_T}{2T} \Theta y_l (c).
\end{array}
$$

Therefore, $\sqrt{H_T} \Delta_{1T} (c) \sim - \Frac{H_T^{3/2}}{2T} \Theta y_l (c),$ and this term is negligible if $H_T / T^{2/3} \rightarrow 0$, if $T\rightarrow \infty$.\vspace{1em}

iii) Effect of $W_{c+u}$\vspace{1em}

Let us now denote $\Delta_{2,T} (c)$ the component associated with the future of the Brownian motion. We have~:

$$
\begin{array}{lcl}
\Delta_{2T} (c) & = & \Frac{1}{H_T} \sum^{H_T}_{h=1} \{ \exp [-\Theta h/T] \Int^{c+h/T}_c \exp (\Theta u) S dW_u \} \\ \\
&=&\Frac{1}{H_T} \sum^{H_T}_{h=1} \sum^h_{k=1} \{ \exp (-\Theta h/T) \Int^{c+k/T}_{c+(k-1)/T} \exp (\Theta u) S dW_u\} \\ \\
&=& \Frac{1}{H_T} \sum^{H_T}_{k=1} \{ [\sum^{H_T}_{h=k} \exp (-\Theta h/T)] \Int^{c+k/T}_{c+(k-1)/T} \exp (\Theta u) S dW_u\} \\ \\
&=&\Frac{1}{H_T} [Id-\exp (-\Theta/T)]^{-1} \\ \\
&&\sum^{H_T}_{k=1} \{ (Id-\exp [-\Theta (H_T-k)/T]) \exp (-\Theta k/T) \Int^{c+k/T}_{c+(k-1)/T} \exp (\Theta u) S dW_u\}.
\end{array}
$$

For expository purpose, i.e. to avoid the matrix notations, let us consider the one-dimensional case. $\Delta_{2T} (c)$ is zero-mean with variance~:

$$
\begin{array}{lcl}
V [\Delta_{2T} (c)] & = & \Frac{s^2}{H^2_T}
                           \Frac{1}{[1-\exp (-\theta/T)]^2}
                           \sum^{H_T}_{k=1} \{ (1-\exp [-\theta (H_T - k)/T])^2 \exp (-2 \theta k/T) \\
                           && \Int^{c+k/T}_{c+(k-1)/T} \exp (2 \theta u) du \}\\ \\
                    & = & \Frac{s^2}{H^2_T}
                          \Frac{\exp (2 \theta c)}{2 \theta [1-\exp (-\theta/T)]^2}
                          \sum^{H_T}_{k=1} [1-\exp (-\theta (H_T-k)/T)]^2 [1-\exp (-2\theta/T)]\\ \\
                    & = & \Frac{s^2}{H^2_T}
                           \Frac{\exp (2 \theta c)}{2 \theta}
                           \Frac{1-\exp (-2\theta/T)}{[1-\exp (-\theta/T)]^2}
                           \sum^{H_T}_{k=1} (1-\exp (-\theta (H_T-k)/T])^2 \\ \\
                    & = & \Frac{s^2}{H^2_T}
                           \Frac{\exp (2 \theta c)}{2 \theta}
                           \Frac{1-\exp (-2\theta/T)}{[1-\exp (-\theta/T)]^2} \\ \\
                    &   &\{H_T - 2
                           \Frac{1-\exp(-\theta H_T/T)}{1-\exp (-\theta/T)} +
                           \Frac{1-\exp (- 2 \theta H_T/T)}{1-\exp (-2 \theta /T)}\}. \;\;\;(*)\end{array}
$$

\noindent Let us now expand the term within brackets when $T\rightarrow \infty, H_T \rightarrow \infty$, $H_T/T \rightarrow 0.$ We get the equivalent expression~:

$$
\begin{array}{l}
  H_T - \frac{2T}{\theta} [ 1 - exp( - \frac{\theta H_T}{T})] + \frac{T}{2 \theta} [ 1 - exp( - \frac{2 \theta H_T}{T})] \\ \\
  \approx  \frac{1}{3} \frac{\theta^2 H_T^3}{T^2}.
\end{array}
$$

\noindent  Therefore, by introducing this expansion in equation $(*)$ we get~:

\begin{eqnarray*}
\lefteqn{V [\sqrt{H_T} \Delta_{2T} (c)]} \\ \\
  & \approx & s^2 \Frac{\exp (2 \theta c)}{2\theta} \Frac{1-\exp (-2\theta/T)}{[1-\exp (-\theta/T)]^2} \frac{1}{3} \frac{\theta^2 H_T^2}{T^2} \\ \\
  & \approx & \frac{1}{3} s^2 \exp (2 \theta c) \frac{H_T^2}{T}.
\end{eqnarray*}

\newpage
     \begin{center}
\textbf{Appendix 4}\vspace{1em}

\textbf{Additional Figures}
\end{center}
\vspace{1em}

\noindent We provide in this appendix additional figures for illustrating  the results of Section 5.2.\vspace{1em}

Figure a.1~: Growth Rates of TFP, Per-Capita GDP, Consumption, Investment and Labor Compensation in US\vspace{1em}

Figure a.2~: ACF and Cross-ACF of the Raw Data\vspace{1em}

Figure a.3~: ACF and Cross-ACF of Filtered ULR component.

%

\newpage

\begin{figure}[h!]
  \centering
  \begin{subfigure}[b]{0.4\linewidth}
    \includegraphics[width=\linewidth,height=5cm]{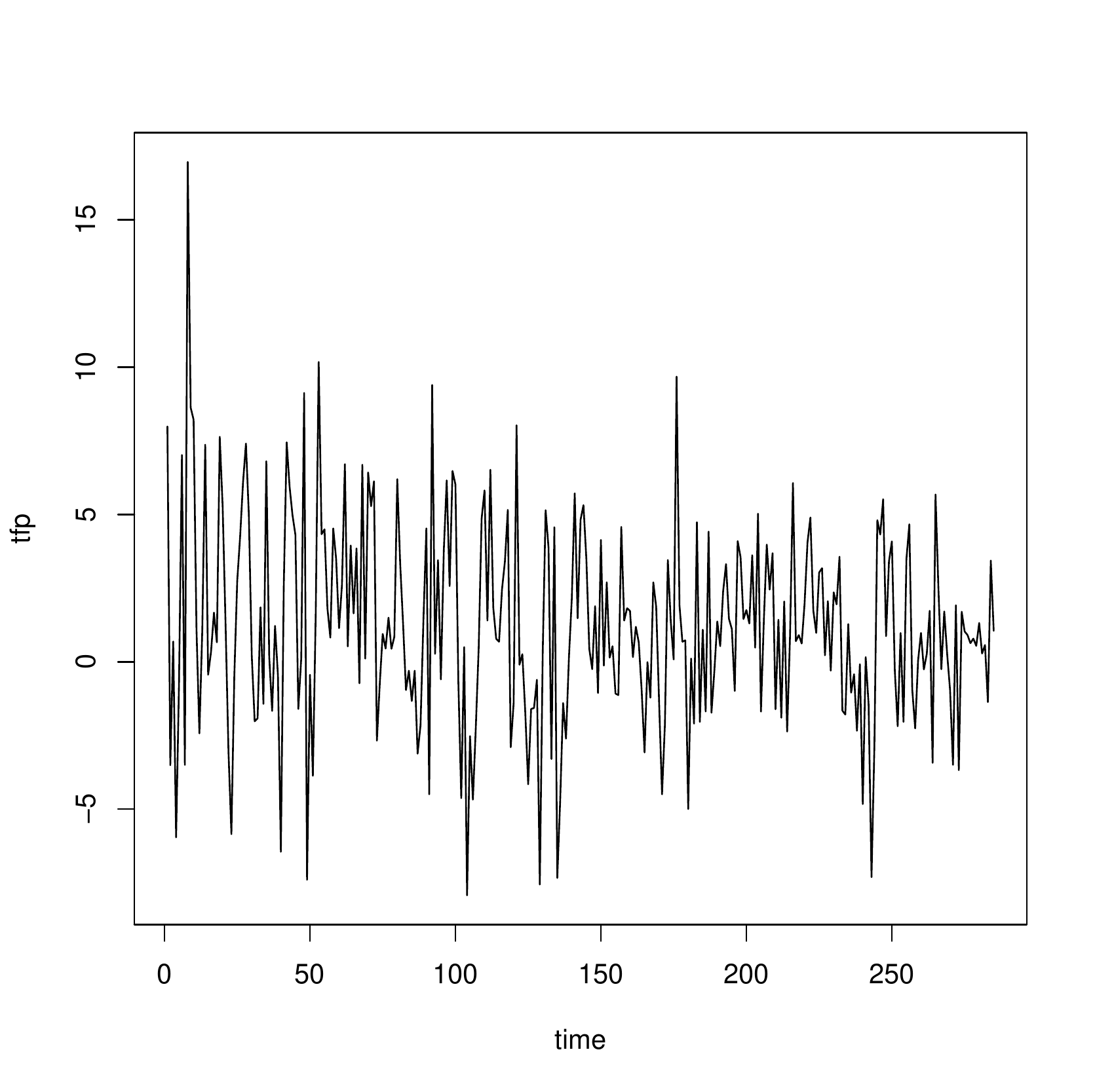}
     \caption{TFP rate}
  \end{subfigure}
  \begin{subfigure}[b]{0.4\linewidth}
\includegraphics[width=\linewidth,height=5cm]{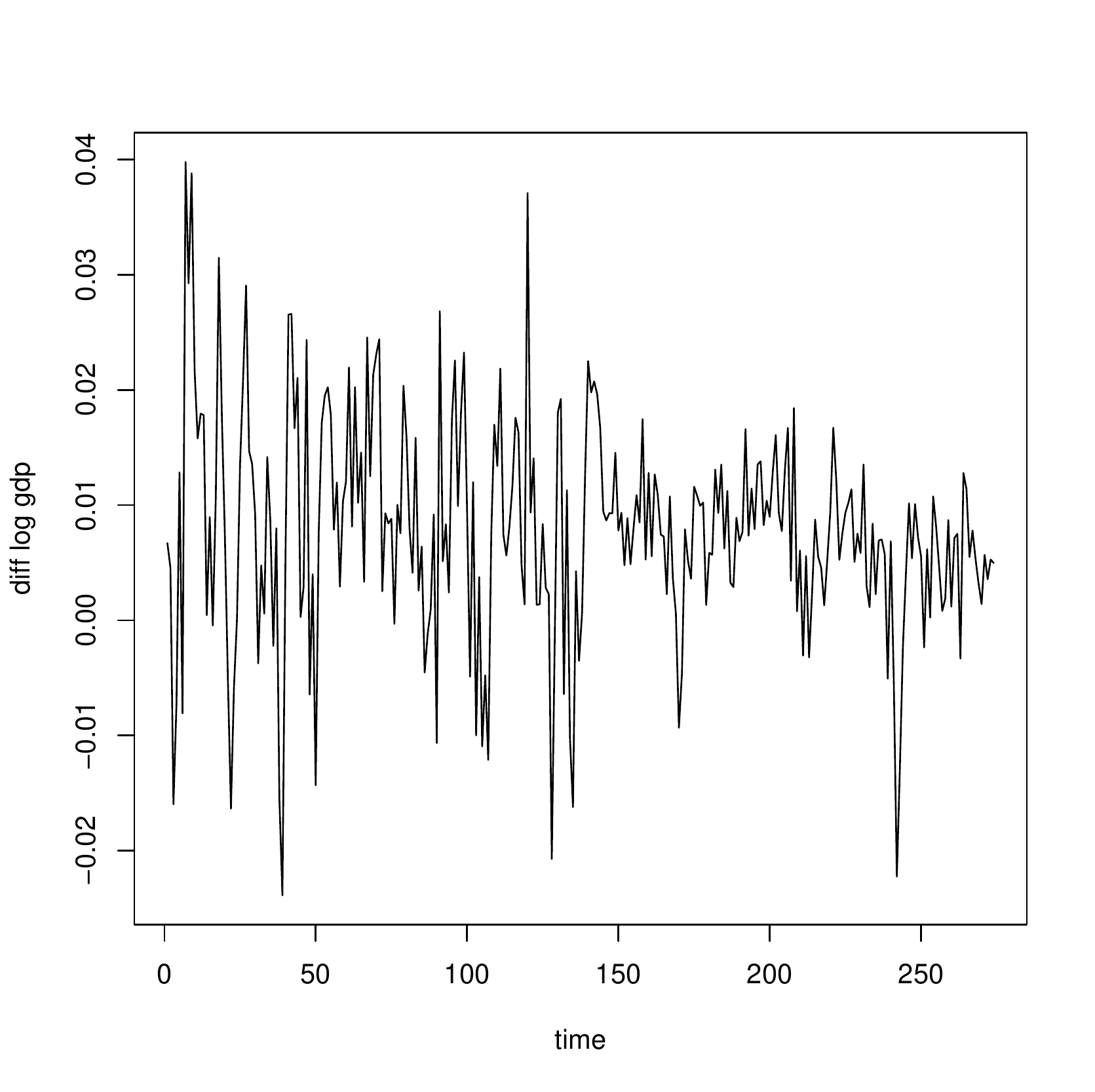}
\centering
    \caption{GDP}
  \end{subfigure}
 \begin{subfigure}[b]{0.4\linewidth}
    \includegraphics[width=\linewidth,height=5cm]{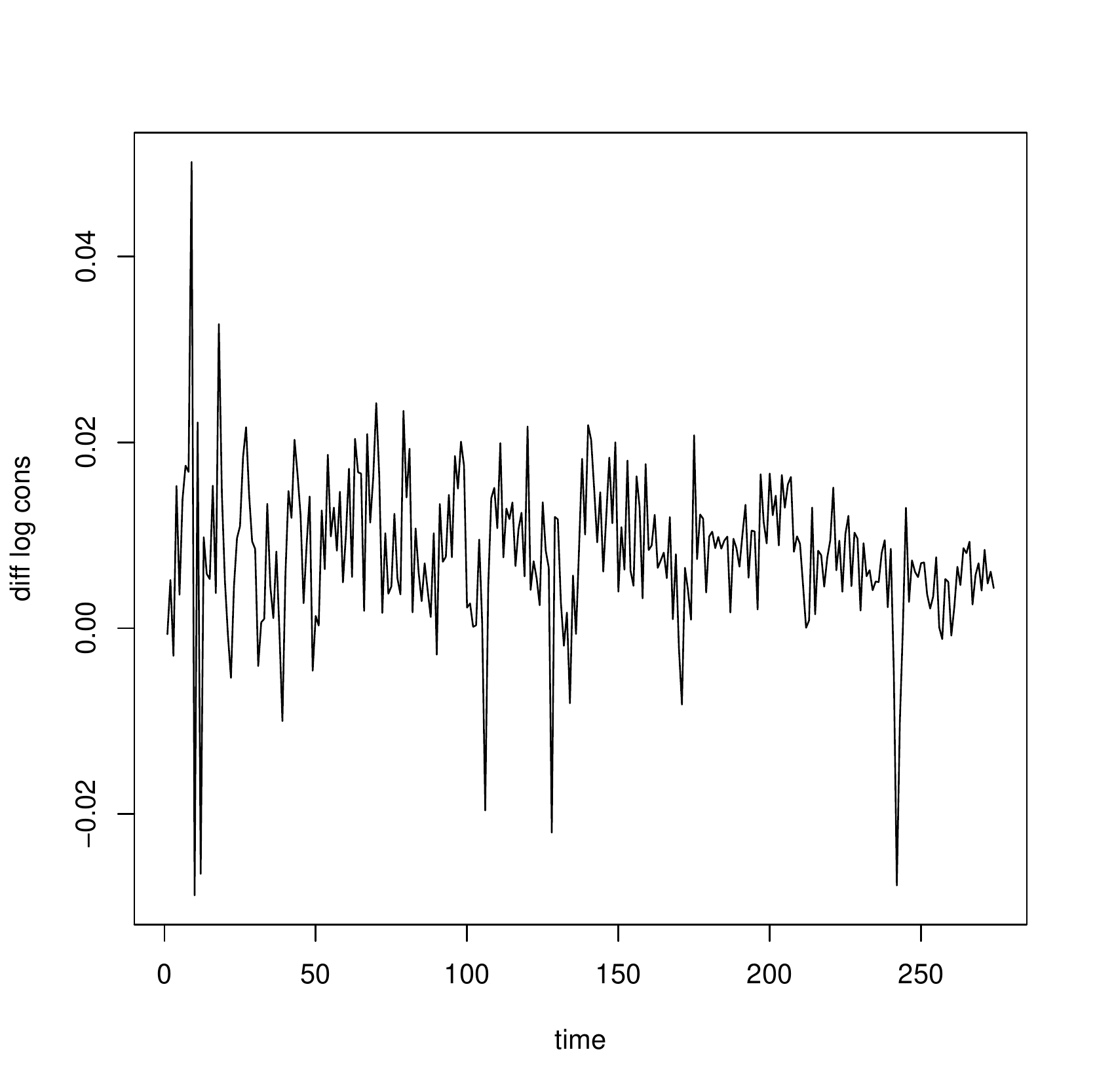}
\centering
    \caption{Consumption}
  \end{subfigure}
 \begin{subfigure}[b]{0.4\linewidth}
    \includegraphics[width=\linewidth,height=5cm]{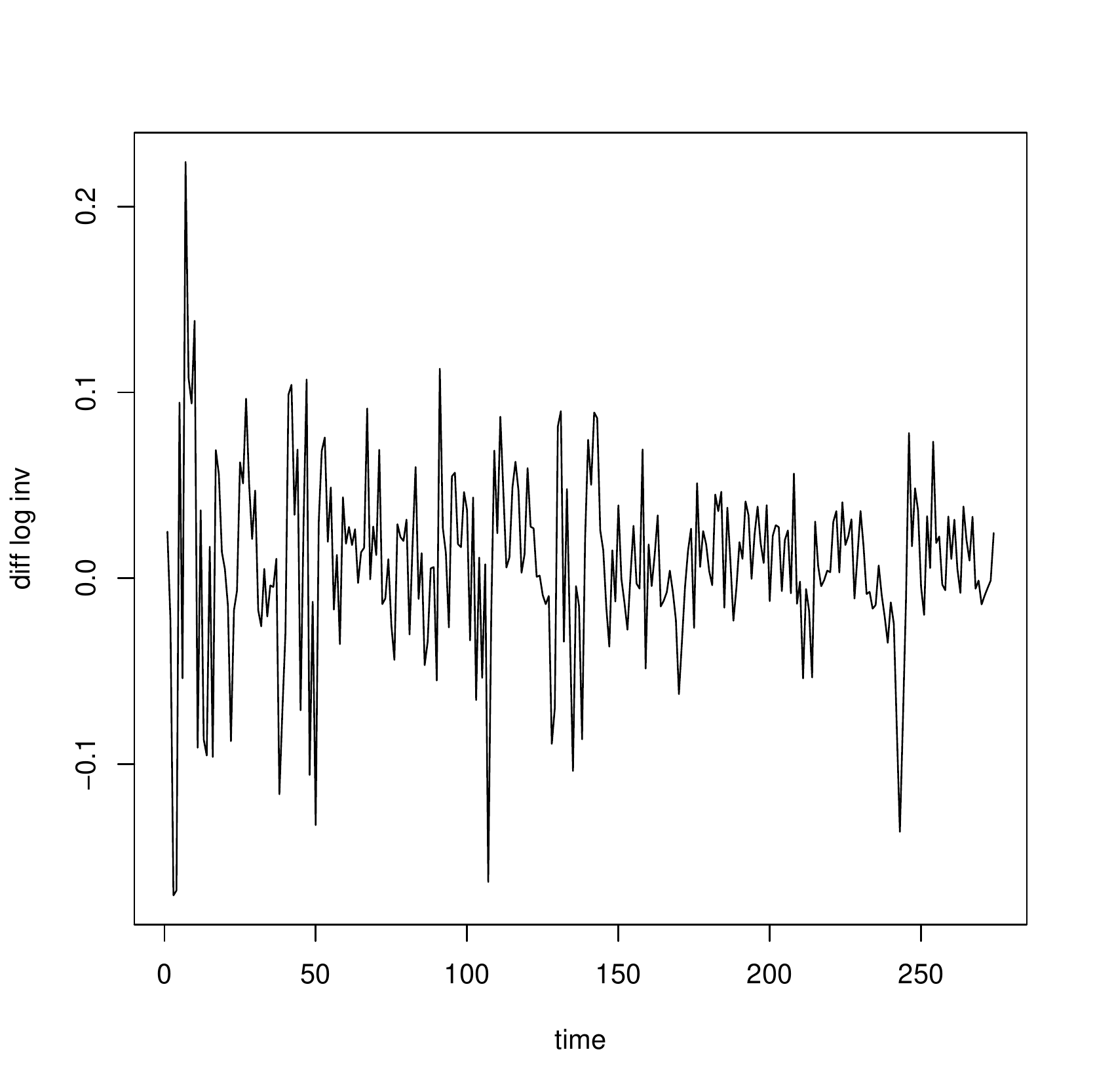}
\centering
    \caption{Investment}
  \end{subfigure}
   \begin{subfigure}[b]{0.4\linewidth}
    \includegraphics[width=\linewidth,height=5cm]{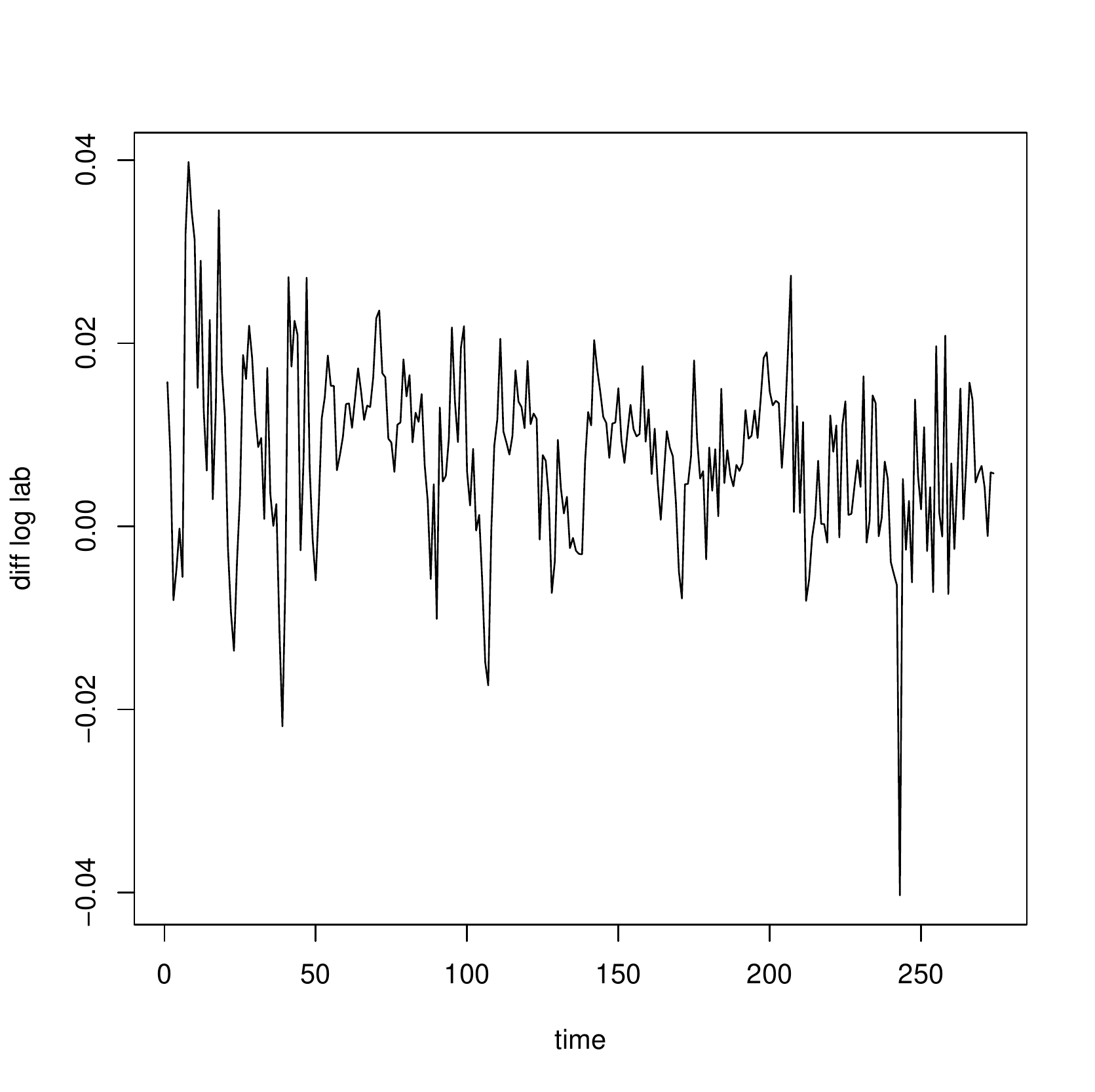}
\centering
    \caption{Labor Compensation}
  \end{subfigure}

\end{figure}
\centerline{Figure a.1 : The series}

\newpage
\centerline{Figure a.2: ACF and Cross-ACF of the Raw Data}\vspace{1em}
\includegraphics[width=\linewidth]{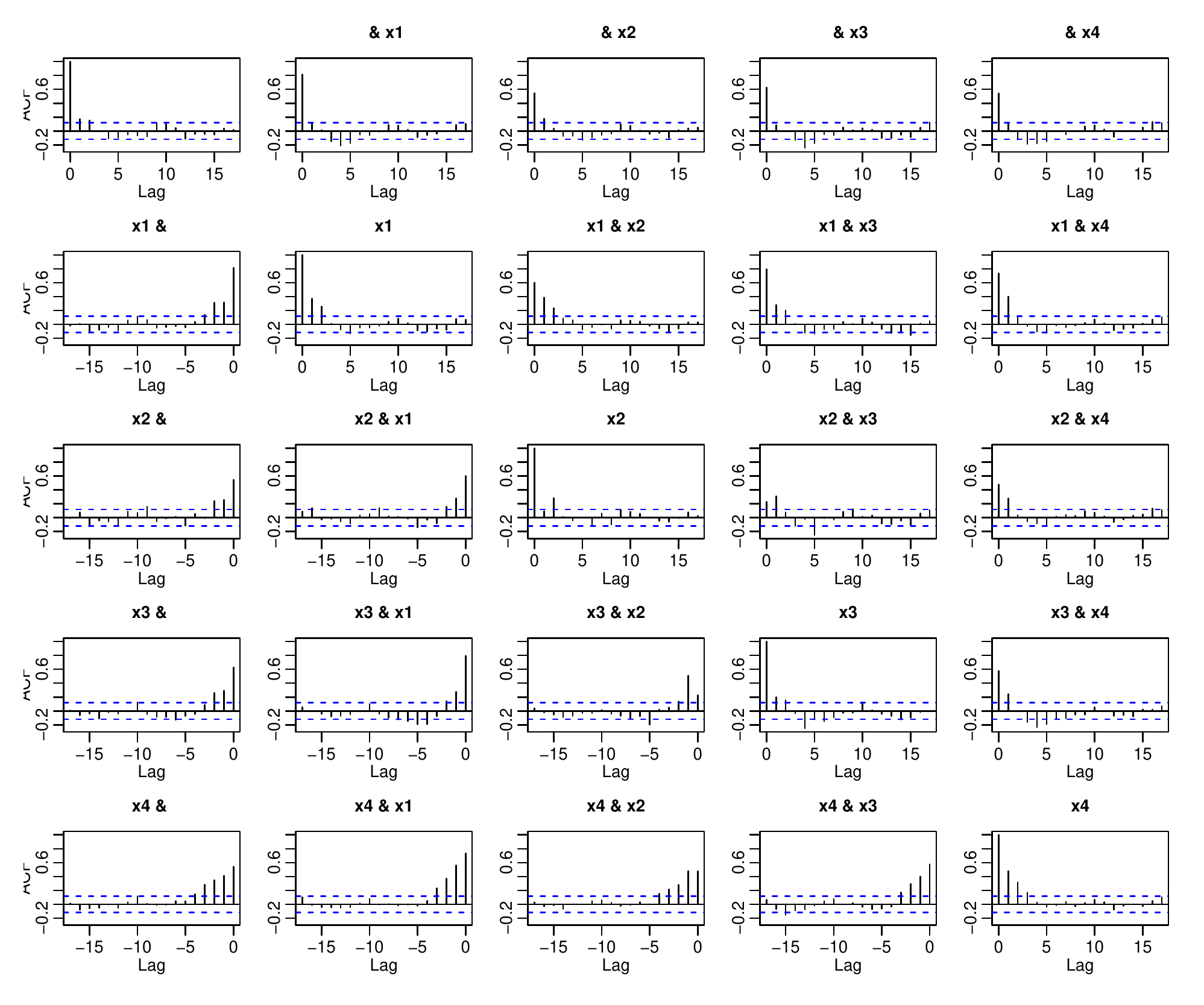}
\newpage

\newpage
\centerline{Figure a.3: ACF and Cross-ACF of Filtered ULR Components}\vspace{1em}
\includegraphics[width=\linewidth]{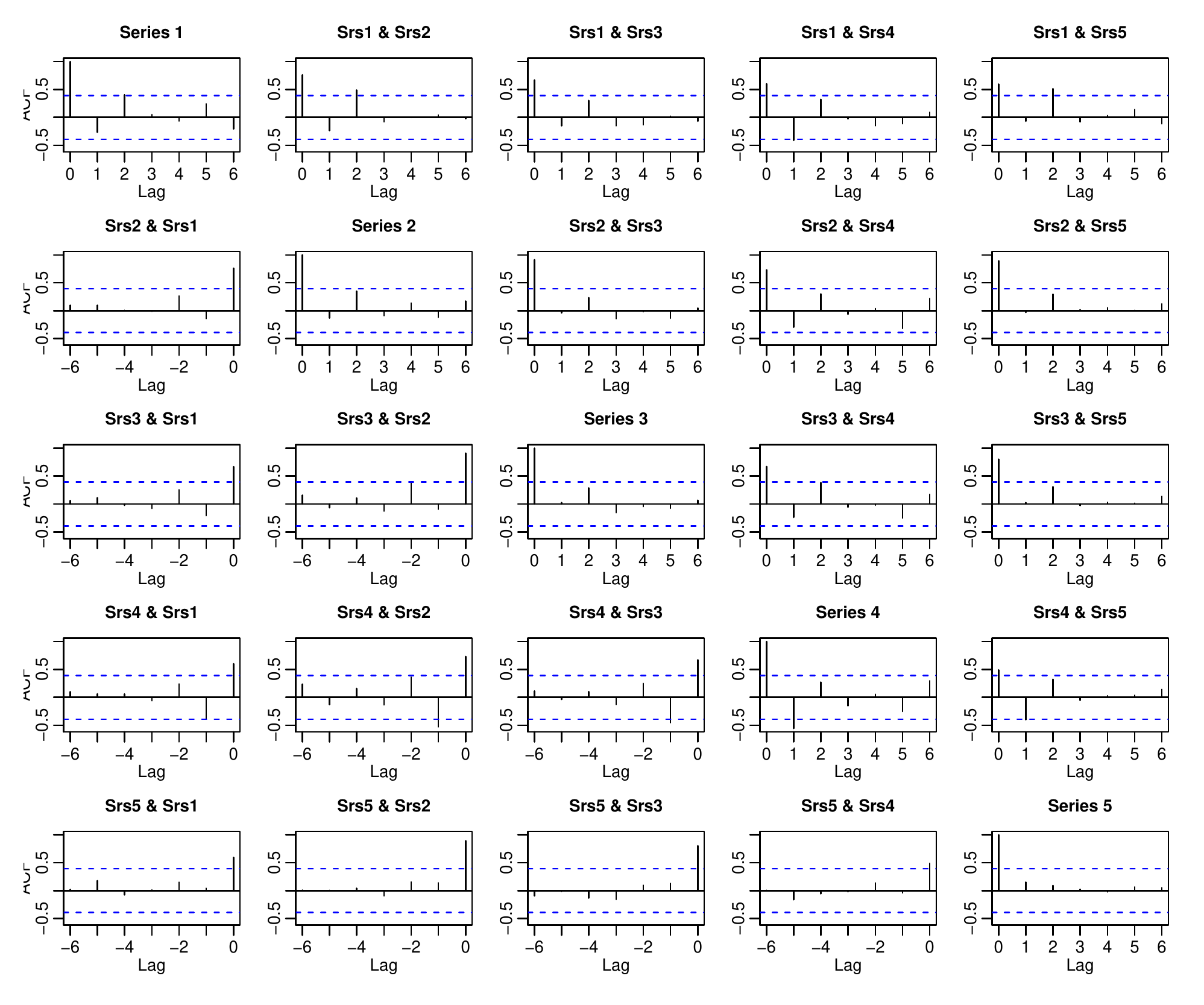}
\newpage

\end{document}